\documentclass[aoas,MSNbibl,nameyear,dvips]{arximspdf}
\usepackage{multirow}
\usepackage{dcolumn}
\usepackage{graphicx}

\doi{10.1214/11-AOAS506}
\volume{6}
\issue{1}
\pubyear{2012}
\firstpage{229}
\lastpage{252}

\makeatletter

\newcolumntype{d}[1]{D{.}{.}{#1}}

\newcommand{\bS}{\mathbf{S}}
\newcommand{\bT}{\mathbf{T}}
\newcommand{\bX}{\mathbf{X}}
\newcommand{\bfb}{{\bolds\beta}}
\newcommand{\bfm}{{\bolds\mu}}
\newcommand{\bfW}{{\bolds\Omega}}
\newcommand{\bfphi}{{\bolds\phi}}
\newcommand{\bflambda}{{\bolds\lambda}}
\newcommand{\ud}{\mathrm{d}}
\newcommand{\uR}{{R}}

\makeatother

\begin{document}
\begin{frontmatter}

\title{Multiple imputation for sharing precise geographies in public use data\thanksref{T1}}
\runtitle{Sharing precise geographies in public use data}

\thankstext{T1}{Supported by NIH Grant R21
AG032458-02. The content is solely the responsibility of the
authors and does not necessarily represent the official views
of the National Institutes of Health.}

\begin{aug}
\author[A]{\fnms{Hao} \snm{Wang}\corref{}\ead[label=e1]{haowang@sc.edu}}
\and
\author[B]{\fnms{Jerome P.} \snm{Reiter}\ead[label=e2]{jerry@stat.duke.edu}}
\runauthor{H. Wang and J. P. Reiter}
\affiliation{University of South Carolina and Duke University}
\address[A]{Department of Statistics\\
University of South Carolina \\
Columbia, South Carolina 29208\\
USA\\
\printead{e1}}
\address[B]{Department of Statistical Science\\
Duke University\\
Durham, North Carolina 27708\\
USA\\
\printead{e2}} %adresu isvedimo komanda gale!
\end{aug}

% HISTORY:
\received{\smonth{8} \syear{2010}}
\revised{\smonth{8} \syear{2011}}

% ABSTRACT
%
\begin{abstract}
When releasing data to the public, data stewards are ethically and
often legally obligated to protect the confidentiality of data
subjects' identities and sensitive attributes. They also strive to
release data that are informative for a wide range of secondary
analyses. Achieving both objectives is particularly challenging when
data stewards seek to release highly resolved geographical information.
We present an approach for protecting the confidentiality of data with
geographic identifiers based on multiple imputation. The basic idea
is to convert geography to latitude and longitude, estimate a bivariate
response model conditional on attributes, and simulate new latitude and
longitude values from these models. We illustrate the proposed methods
using data describing causes of death in Durham, North Carolina. In the
context of the application, we present a straightforward tool for
generating simulated geographies and attributes based on regression
trees, and we present methods for assessing disclosure risks with such
simulated data.
\end{abstract}

% KEYWORDS
%

\end{frontmatter}

%s1 #&#
\section{Introduction}
Statistical agencies, research centers and individual researchers
frequently collect geographic data as an integral part of their
studies. Geographic data can be highly beneficial for analyses. In
studies of aging, for example, they can reveal areas where elderly
people live in high densities, which is useful for policy and
planning; they can illuminate how environmental factors impact the
health and quality of life of elderly people; and, through
contextual data, they can yield insights into the social and
economic conditions and lifestyle choices of the elderly. Analysts
who do not account for spatial dependencies may miss important
geographic trends and differences, potentially resulting in invalid
inferences.

Geographic variables also are among the most challenging data to
share when making a primary data source available to other
researchers and the broader public. Very fine geography, while
facilitating detailed spatial analyses, enables ill-intentioned
users to infer the identities of individuals
in the shared file. Even modestly coarse geography can be risky in
the presence of demographic or other readily available attributes,
which when combined may identify individuals in the shared file.
Such identifications are problematic for data collectors, who are
ethically and often legally obligated to protect data subjects'
confidentiality. To reduce the risks of disclosures, data collectors
typically delete or aggregate geographies to high levels before
sharing data. Unfortunately, deletion and aggregation sacrifice the
quality of analyses that utilize finer geographic detail.
%The difficulty of sharing data with geographic identifiers
%led a recent U.S. National Research Council panel on protecting
%confidentiality with geographic data to conclude that, ``At this
%time, no known technical strategy or combination of technical
%strategies for managing linked spatial-social data adequately
%resolves conflicts among the objectives of data linkage, open
%access, data quality, and confidentiality protection across datasets
%and data uses'' \citet[][p. 2]{nrc07}.

We propose to protect the confidentiality of data with fine geographic
identifiers by simulating values of geographies and other identifying
attributes from statistical models that capture the spatial
dependencies among the variables in the collected data. These simulated
values replace the collected ones when sharing data. To enable
estimation of variances, the data steward generates several versions of
the data sets for dissemination, resulting in multiply-imputed,
partially synthetic data sets [\citet{little1993},
\citet{reiterpartsyn}]. Such
data sets can protect confidentiality, since identification of units
and their sensitive data can be difficult when the geographies and
other quasi-identifiers in the released data are not actual, collected
values. And, when the simulation models faithfully reflect the
relationships in the collected data, the shared data can preserve
spatial associations, avoid ecological inference problems, and
facilitate small area estimation.

%Our approach takes advantage of recent advances in spatial modeling,
%%improvements in converting addresses into numerical values (and
%vice-versa) amenable to modeling, and the latest methods for
%evaluating data confidentiality and data quality. Utilizing these
%techniques gives us the potential to preserve spatial relationships
%at levels unattainable with existing methods of data protection.

The remainder of the article is as follows. In Section \ref{sec2} we
describe some of the shortcomings of current approaches to
protecting data with geographies, and we motivate the use of
multiple imputation for releasing public use data with highly
resolved geographies.
%In
%Section 3, we describe the steps for generating synthetic
%geographies. SEPARATE SECTION? OR PUT IN SECTION 2?
%straightforward data simulator based on sequential regression trees
%that can be adapted to synthesize geographies or attributes. In
%In Section 4, we describe a framework for assessing disclosure risks in
%the synthetic data.
In Section \ref{sec3} we generate multiply-imputed, partially synthetic versions
of a spatially-referenced data set
describing causes of death in Durham, North Carolina. As part of the
application, we present an easy-to-implement data
simulator based on sequential regression trees for
synthesizing highly-resolved geographies or attributes. We also
describe methods
for assessing disclosure risks for data with synthetic geographies.
These include (i) a new measure for quantifying the
risks that the original geographies could be recovered from the
simulated data, and (ii) a measure for
assessing risks of re-identifications based on the approach of
\citet{reitermitra07}. In Section \ref{discussion}
we conclude with issues for implementation of the approach.

%s2 #&#
\section{Motivation for using simulated geographies}\label{sec2}

At first glance, releasing or sharing safe data seems a
straightforward task: simply strip unique identifiers like names and
tax identification numbers before releasing data. However,
these\vadjust{\goodbreak}
actions alone may not suffice when other readily available
variables, such as geographic or demographic data, remain on the
file. These quasi-identifiers
can be used to match units in the released data to other databases.
When the quasi-identifiers include geographic variables, the risks
of identification disclosures can be extremely high. For example,
Sweeney [(\citeyear{sweeney2001}), pages 51 and 52] showed that 97\% of
the records in
a publicly available voter registration list for Cambridge, MA,
could be identified using only birth date and 9-digit zip code.
Because of the disclosive nature of geography, the U.S. Health
Insurance Portability and Accountability Act (HIPAA) Privacy Rule
requires that, when sharing certain health data, the released
geographic units comprise at least 20,000 people [\citet{hipaa},
page~82543].

Data stewards can protect confidentiality by restricting public
access to the data. For example, analysts can use the data only in
secure data enclaves, such as the Research Data Centers operated by
the U.S. Census Bureau. Or, analysts can submit queries to remote
access systems that provide statistical output without
revealing the data that generated the output. While useful,
restricted access strategies are only a partial solution. Analysts
who do not live near a secure data enclave, or do not have the
resources to relocate temporarily to be near one, are shut out from
this form of access. Gaining restricted access can require months of
proposal preparation and background checks; analysts cannot simply
walk in to any secure data enclave and immediately start working
with the data. Remote access servers limit the scope of analyses
and details of output, since clever queries can reveal individual
data values [\citet{statsciserverkarr}]. Performing exploratory data
analysis and checking model fit are difficult without
access to record-level data. Hence, as recommended by two recent
National Research Council panels on data confidentiality, to
maintain the benefits of wide dissemination, it is necessary to
supplement restricted access strategies with readily available,
record-level data [National Research Council (\citeyear{nrc05,nrc07})].

%s2.1 #&#
\subsection{Common approaches to protecting geography}\label{ExistAppr}

Data stewards commonly employ several strategies for protecting
confidentiality when sharing data with geographic identifiers.
However, these methods can have serious impacts on the quality of
the released data, as we now describe.

\subsubsection*{Data suppression} Data stewards can suppress
geography or attributes from data releases. The intensity of
suppression can range from not releasing entire variables, for
example, stripping the file of all geographic identifiers, to not
releasing small subsets of values, for example, blanking out
sensitive attribute values. An example of the former is the Health
and Retirement Study: the public use data do not contain any
geographic information on relocations [\citet{hrs}, page 14].
Increasing the intensity of suppression generally increases data
protection and decreases data quality.\vadjust{\goodbreak} While intense suppression can
reduce risks, it has repercussions for inferences. Wholesale
deletion of geographic identifiers disables any spatial analysis.
When relationships depend on the omitted geography, analysts'
inferences are biased. Selective suppression of geography or
attributes creates data that are missing
not at random, which complicates
analyses for users. When there are many records at
risk, as is likely the case when the data have fine geographic
identifiers, data stewards may need to suppress so many values to
achieve satisfactory protection that the released data have very
limited quality for spatial analysis.\vspace*{-3pt}

\subsubsection*{Data aggregation} Data stewards can coarsen geography or
other variables, for example, releasing addresses at the block or
county rather than parcel level, or releasing ages in five year
intervals. Aggregation reduces disclosure risks by turning unique
records---which generally are most at risk---into nonunique
records. For example, there may be only one person with a
particular combination of demographic characteristics in a street
block, but many people with those characteristics in a state.
Releasing data for this person with geography at the street level
might have a high disclosure risk, whereas releasing the data at the
state level might not. The amount of
aggregation needed to protect confidentiality depends on the nature
of the data. When other identifying attributes are present, such as
demographic characteristics, high-level aggregation of the
geographic identifiers may be needed to achieve adequate protection.
For example, there may be only one person of a certain age, sex,
race and marital status---which may be available to ill-intentioned
users at low cost---in a particular county, so that coarsening
geographies to the county level provides no greater protection for
that person than does releasing the exact address.

Aggregation preserves analyses at the level of aggregation. However,
it can create ecological
inference fallacies [\citet{ecologicalinference},
\citet{freedman04}] at lower levels of aggregation.
% i.e., relationships observed at high levels of
%aggregation do not exist at lower levels.
Additionally, when
geography is highly aggregated, analysts may be unable to detect
important local spatial dependencies. Despite these limitations,
aggregation is the most widely used solution to protect data with
geographic identifiers and is routinely implemented by government
agencies and other data collectors. The U.S. Census Bureau,
for example, does not release geographic identifiers below
aggregates of at least 100,000 people in public use files of census
data. The public use files for the Health and Retirement Study
aggregate geography to ``a level no higher than U.S. Census Region
and Division'' [\citet{hrs}, page 14].

Aggregation also is frequently used to disguise values in the tails
of nongeographic quasi-identifiers, especially age. The HIPAA
requires that all ages above 89 be aggregated into and shared as a
single category, ``90 or older.''\vspace*{-3pt}

\subsubsection*{Random noise addition} Data stewards can disguise
geographic and other attribute values by adding some randomly selected
amount to each confidential observed value. For geographic
attributes,\vadjust{\goodbreak}
this involves moving an observed location to another randomly drawn
location, usually within a~circle of some radius $r$ centered at the
original location. The quality of inferences and the amount of
protection depend crucially on $r$. When a large $r$ is needed to
protect confidentiality---as is likely the case when data contain
readily available quasi-identifiers---inferences involving spatial
relationships can be seriously degraded [\citet{rushton},
\citet{guttman}]. Adding random noise to attribute values
introduces measurement error, which inflates variances and attenuates
regression coefficients [\citet{fuller1993}].

\subsubsection*{Random data swapping} Data stewards can swap data values
for selected records, for example, switch values of age, race and sex for
at-risk records with those for other records, to discourage users
from matching, since matches may be based on incorrect data
[\citet{dalenius}, \citet{fienmac}]. Swapping is used
extensively by government
agencies. It is generally presumed
that swapping fractions are low---agencies do not reveal the rates
to the public---because swapping at high levels destroys
relationships involving the swapped and unswapped variables. Because
data stewards might have to swap all geographic identifiers to ensure
released records do not have their actual geographies, swapping is
not effective for highly resolved geographic identifiers.

%s2.2 #&#
\subsection{Proposed approach: Simulate geographic identifiers}
\label{MultiImput}
The main limitation of the approaches in Section \ref{ExistAppr} is that
they perturb the geography or other quasi-identifiers with minimal
or no consideration of the relationships among the variables. Our
proposed approach explicitly aims to preserve relationships among
the geographic and other attributes through statistical modeling. At
the same time, replacing geographic and other quasi-identifiers with
imputations makes it difficult for ill-intentioned users to know the
original values of those variables, which reduces the chance of
disclosures.

Our approach differs from the recent proposal of
\citet{ZhouEtAl2010}, who use spatial smoothing to mask nongeographic
attributes at the original locations.
%rather, they
%mask nongeographic attributes at fixed locations.
Releasing the original locations can result in high risks of
identification disclosures when the data include fine geography.
\citet{ZhouEtAl2010} do not intend to deal with these risks, whereas
we explicitly seek to do so. We note that spatial smoothing could
be used to mask attribute values after synthesis of locations.

To illustrate how our approach might work in practice, we modify
the setting described by \citet{reiterchance}. Suppose that a
statistical agency has collected data on a random sample of 10,000
heads of households in a state. The data comprise each person's
street block, age, sex, income and an indicator of disease status.
Suppose that combining street block, age and sex uniquely
determines a large percentage of records in the sample and the
population.\vadjust{\goodbreak} Therefore, the agency wants to replace street block,
age and sex for all people in the sample---or possibly only a
fraction of the three variables, for example, only street block for some
records and only age and sex for others---to disguise their
identities. The agency generates values of street block, age and
sex for these people by randomly simulating values from their joint
distribution (see Section \ref{SynModels}), conditional on their
disease status and income values.
This distribution is estimated with the collected data. The result
is one partially synthetic data set. The agency repeats this process,
say, ten times, and these ten data sets are released to the public.
% for
%secondary analysis.
%The multiple datasets enable proper variance estimation.

To illustrate how a secondary data analyst might utilize these
shared data sets, suppose that the analyst seeks to fit a logistic
regression of disease status on income, age, sex and indicator
variables for the person's county (obtained by aggregating the
released, simulated street blocks). The analyst first estimates the
regression coefficients and their variances separately in each
simulated data set using standard likelihood-based estimates and
standard software. Then, the analyst averages the estimated
coefficients and variances across the simulated data sets. These
averages are used to form 95\% confidence intervals based on the
simple formulas developed by \citet{reiterpartsyn}, described below.

The agency creates $m$ partially synthetic data sets, $D^{(1)},
\ldots, D^{(m)}$, that it shares with the public. Let $Q$ be the
secondary analyst's estimand of interest, such as a regression
coefficient or population average. For $l = 1, \ldots, m$, let
$q^{(l)}$ and $u^{(l)}$ be respectively the estimate of $Q$ and the
estimate of the variance of $q^{(l)}$ in synthetic data set
$D^{(l)}$. Secondary analysts use $\bar{q}_{m} = \sum_{l=1}^{m}
q^{(l)}/m$ to estimate $Q$ and $T_{m} = \bar{u}_{m}+b_m/m$ to
estimate $\operatorname{var}(\bar{q}_{m})$, where $b_m = \sum_{l=1}^{m}
(q^{(l)} - \bar{q}_{m})^{2} / (m-1)$ and $\bar{u}_{m} =
\sum_{l=1}^{m} u^{(l)}/m$. For large samples, inferences for $Q$ are
obtained from the $t$-distribution, $(\bar{q}_m - Q) \sim
t_{\nu_m}(0, T_m)$, where the degrees of freedom is ${\nu_m} = (m -
1) [1+ m\bar{u}_{m}/b_m]^{2}$. Details of the derivations
of these methods are in \citet{reiterpartsyn}. Tests of significance
for multicomponent null hypotheses are derived by
\citet{reitermulti}.

%When the street block, age, and sex values are simulated from their
%true probability distribution, the simulated data should have
%similar characteristics on average as the collected data.
%There is
%an analogy here to random sampling. Some true distribution of the
%variables exists in the population. The collected data are just a
%random sample from that population distribution. If the agency
%simulates data from that same distribution---which is guaranteed for
%the bivariate distribution of income and disease status in this
%example, since they remain unchanged---it essentially creates
%different random samples from the population.
%Hence, the analyst using these partially simulated datasets
%essentially analyzes
%alternative samples from the population.
%The on average caveat is important: parameter estimates from any one
%simulated dataset are unlikely to equal exactly those from the
%collected data. The simulated parameter estimates are subject to
%two sources of variation, namely (i) sampling the collected data and
%(ii) generating simulated values. It is not possible to estimate
%both sources of variation from only one released simulated dataset.
%However, it is possible to do so from multiple simulated datasets,
%which explains why the multiple imputation framework applies.

Partially synthetic data sets can have positive data utility
features. When data are simulated from distributions that reflect
the distributions of the collected data,
Reiter (\citeyear{reiterpartsyn,reitermi,reitermulti})
shows that analysts can obtain valid
inferences (e.g., 95\% confidence
intervals contain the true values 95\% of the time) for wide classes
of estimands. These inferences are determined by combining standard
likelihood-based or survey-weighted estimates; the analyst need not
learn new statistical methods or software to adjust for the effects
of the disclosure limitation. The released data can include
simulated values in the tails of distributions, for example, there is no
top-coding of ages or incomes [however, it is challenging to develop
synthesis models that simultaneously protect confidentiality and
preserve inferences when data are very sparse in tails; see
\citet{reiter2002a}]. Because many quasi-identifiers
including geography can be simulated, finer details of geography can
be released, facilitating estimation for small areas and spatial
analyses.

There is a cost to these benefits: the validity of
inferences depends on the validity of the models used to generate
the simulated data. The extent of this dependence is driven by the
nature of the synthesis. For example, when all of age and sex are
synthesized, analyses involving those variables reflect only the
relationships included in the data generation models. When the
models fail to reflect certain relationships accurately, analysts'
inferences also will not reflect those relationships. Similarly,
incorrect distributional assumptions built into the models will be
passed on to the users' analyses. On the other hand, when replacing
only a select fraction of age and sex and leaving many original
values on the file, inferences are less sensitive to the assumptions
of the simulated data models. In practice, this dependence means
that data stewards should release information that helps analysts decide
whether or not the simulated data are reliable for their analyses.
For example, data stewards might include the data generation models
(without parameter estimates) as attachments to public releases of
data. Or, they might include generic statements that describe the
imputation models, such as ``Main effects and interactions for age,
sex, income and disease status are included in the imputation
models for street blocks.'' Analysts who desire finer detail than
afforded by the imputations may have to apply for restricted access
to the collected data.

%Describing the imputation models is necessary, but it is not
%sufficient. Agencies also should release simulated data generated
%from the models. Some analysts are not able to generate simulated
%data given the models; they need agencies to do it for them. Even
%when analysts can do so, it is a cumbersome burden to place on them.
%Analysts may desire some function of the simulated data that is hard
%to estimate from the model parameters, but easy to determine from
%the simulated data. Furthermore, releasing exact details about the
%synthesis models including parameter estimates can increase
%disclosure risks \citet{reitermitra07}.

When generating partially synthetic data, the data steward must choose which
values to synthesize and must specify models to simulate replacements
of those values. In most existing partially synthetic data sets,
stewards replace all values of variables that they deem to be either
(i) readily available to ill-intentioned users seeking to
identify released records, or (ii) too sensitive to risk releasing
exactly. However, it
may be sufficient from a confidentiality perspective to replace only
portions of some variables; see \citet{litragliu}. The process of
specifying synthesis
models is typically iterative: the data steward creates synthetic
data using a posited model, checks the quality of a large number of
representative analyses with the synthetic data, and
adjusts the models as necessary to improve quality while maintaining
confidentiality protection. For examples of this process, see
\citet{reitdrechcensus} and \citet{lbdisr}.

The data steward also must determine $m$, that is, how many synthetic
data sets to release. Generally, increasing $m$ results in decreased
standard errors in secondary analyses. However, increasing $m$
results in greater data storage costs and
possibly increased disclosure risks [\citet{reitermitra07}]. When
small fractions of values are synthesized (e.g., around 10\%), the
efficiency gains
from increasing $m$ are typically modest, so that data stewards can
make $m$ modest, for example, $m=5$, to keep risks and storage costs
comparatively low. When large fractions of values are replaced,
efficiency gains from increasing $m$ can be substantial
[\citet{reitdrechcensus}]. In such cases, we recommend that data stewards
select the largest $m$ that still offers acceptable risks and storage costs.

%%%%%%MAKE SUBSECTION?
%s2.3 #&#
\subsection{Synthesis models for sharing precise geographies}
\label{SynModels}

Our strategy for simulating geographies involves four general steps.
First, the data steward converts the geographic variables on the file to
latitudes and longitudes (possibly, using UTM projection to Eastings
and Northings). When the collected geographies are aggregated rather
than precise locations, the data steward uses a typical value for the
location of all records in that area; for example, use the latitude
and longitude of the centroid of the street block. Second, the
data steward estimates a model for latitudes and longitudes conditional on
other variables in the data set.
%We call this the {\em simulationmodel}.
Third, using this model, the data steward simulates new
latitudes and longitudes for every record in the file.
%As we shall describe, the data steward's model for generating new
%latitudes and
%longitudes might differ from the estimation model; hence, we call it
%the {\em
% simulation model}.
Fourth and finally, the data steward releases multiple draws of the
simulated latitudes and longitudes along with the other
attributes---which also might be altered to protect
confidentiality, for example, \citet{ZhouEtAl2010}---in the
original file.

%%%%%%%%%%% MOVE THIS STUFF TO SUPPLEMENT?
% Alternatively, to avoid
%public perceptions that the simulated latitudes and longitudes
%represent specific sampled locations, the data steward can release the

%In some settings, the data steward may wish to release the
%%locations slightly, for example to ensure that
%synthetic geographies in some slightly aggregated form, such as tax
%parcels or street blocks, rather than as latitudes and longitudes.
%This may be desired, for example, by secondary analysts of the data.
%The steward can use reverse geocoding to determine the aggregated
%value that corresponds to the simulated latitude and longitude; for
%example, find the closest parcel/block for a simulated
%latitude and longitude value. The aggregated value can be obtained
%either from a map of the area, including possibly nonsampled
%aggregates, or from only the
%aggregates of the sampled records.

We expect that, in general, some attributes in the data will exhibit
spatial dependence. When considering location as the response
variable, this implies a joint distribution for latitude and
longitude that depends on the attributes and is possibly
multi-modal.
%This stems form the fact that some attributes in the
%dataset will exhibit spatial dependence whence, upon considering
%location as the response surface, will result in multi-modality.
For example, people of similar age, socio-economic status and other
demographic characteristics tend to cluster in neighborhoods, and
certain demographic characteristics may be highly prevalent in
multiple locations but absent in others. If we ignore these features
when simulating geographies---or alter geography with approaches
that do not explicitly account for these associations---the spatial
relationships in the data will be altered or destroyed.

To illustrate some possible response models for locations,
let $\phi_i$ and $\lambda_i$ denote the latitude and longitude,
respectively, for data subject $i$. Let $\mathbf{x}_i$ denote the
$p$ nongeographical attributes for data subject $i$. One family of
convenient response models is
$(\lambda_i, \phi_i) \sim N(\bfm_i, \bfW_i)$,
where each $\bfm_i = h(\mathbf{x}_i)$ is a $2 \times1$ vector of
unknown means, $h(\mathbf{x}_i)$ is a function of the covariates,
and each $\bfW_i$ is an unknown $2 \times2$ covariance matrix. A
simple implementation is a~bivariate regression model with
$h(\mathbf{x}_i) = \beta_0 + \sum_{j=1}^p h_j(x_{ij})\beta_j$, where
each~$h_j$ is a spline for variable $j$ and $\bfW_i = \bfW$ for all
$i$. An alternative is a mixture model with $h(\mathbf{x}_i) =
\beta_{i0} + \sum_j h_j(x_{ij})\beta_{ij}$, where
%$\beta_i$ and
%$\Lambda_i$ come from one of $K$ bivariate normal distributions.
$\bfb_i = (\beta_{i0},\ldots,\beta_{ip}) $ and $\bfW_i$ come from
$K$ mixture components.

In specifying a response model for locations, the data steward
should include components of $\mathbf{x}$ that vary with spatial
locations. The data steward also should seek a flexible model that can
adapt to a potentially complex response distribution. In the
application, we describe a semi-automated approach for approximating
the response distribution that can be easily implemented by data
stewards. We emphasize, however, that the idea of treating latitude
and longitude as a response is general, and that data stewards
can improve the quality of the released data by tailoring the
response model to their particular problem.

To our knowledge, treating geography as a continuous response and
releasing simulated draws from its distribution
has not been previously implemented. However, partially
synthetic data are used to protect locations
in the Census Bureau's OnTheMap project [\citet{onthemap}]. In
that project, \citet{onthemap} synthesize the street blocks where
people live conditional on the street blocks where they work and other
block-level attributes. They use multinomial regressions to simulate
home-block values, constraining the possible outcome space for each
individual based on where they work. Our approach
differs from the OnTheMap modeling in that (i) we model more
precise geography, that is, continuous versions of latitudes and
longitudes, than discrete street blocks, and (ii) we do not rely on
a fixed set of geographic locations, that is, where people work, to
anchor the synthesis models. Furthermore, for settings
with high-dimensional $\mathbf{x}_i$ and
no obvious way to set constraints on the outcome
space, multinomial regressions can be computationally
demanding if even estimable, whereas continuous response models are
readily estimated.

%
%t1 #&#
\begin{table}[b]
\tablewidth=260pt
\caption{Description of variables used in the empirical
study} \label{tabdescription}
\begin{tabular*}{\tablewidth}{@{\extracolsep{\fill}}ll@{}}
\hline
\textbf{Variable} & \multicolumn{1}{c@{}}{\textbf{Range}}\\
\hline
Longitude & Recoded to go from 1--100 \\
Latitude & Recoded to go from 1--100 \\
Sex & Male, female\\
Race & White, black\\
Age (years) & 16--99\\
Autopsy performed & Yes, no, missing\\
Autopsy findings & Yes, no missing\\
Marital status & 5 categories\\
Attendant & Physician, medical
examiner, coroner
\\
Hispanic & 7 categories\\
Education (years) & 0--17 years\\
Hospital type & 8 categories \\
\hline
\end{tabular*}
\end{table}

%s3 #&#
\section{Application: Protecting a cause of death file}\label{sec3}

We now apply the multiple imputation approach to create
disclosure-protected data on a subset of North Carolina (NC)
mortality records in 2002. The data include precise longitudes and
latitudes of
deceased individuals' residences, as well as a variety of variables
related to manner of death; we consider the subset of variables in
Table \ref{tabdescription}. These mortality data are in fact publicly
available and so do not require disclosure protection. Nonetheless,
they are ideal test data for methods that protect confidentiality of
geographies since, unlike many data sets on human individuals, actual
locations are available and can be revealed for comparisons. Access
to the data is managed by the Children's Environmental Health
Initiative at Duke University per agreement with the state of NC.

%The complete NC data comprises about 288635 units.
%The data comprise variables related to manner of death;
We use individuals whose place of residence was one of seven
contiguous postal zones in Durham, NC.
%27701, 27703, 27704, 27705, 27707, 27712 and 27713.
These areas are heterogeneous in terms of population density and
characteristics.
%with communities ranging from the urban area of downtown Durham, to
%the area that are close to research triangle park, and to the area
%that are close to Eno River State Park.
%We do not reveal the actual year and postal zones to help safeguard
%confidentiality, since we present maps based on genuine data in
%Section 3.4.
For simplicity, we include only individuals with race of black and
white---which comprised $99\%$ of all records in these postal
zones---resulting in $n=2\mbox{,}670$ observed cases. We also collapse the
cause of death
variable into two levels: death from diseases of the circulatory and
respiratory system, and death from all other causes. We consider
this binary variable, which we label $Y$, as the outcome for
regression models.

% The generation of synthetic datasets involves variables that are
% displayed in Table \ref{tabdescription}. Race has 10 categories in
% the original data with white and black compromising of $99\%$ of
% all records. For utility evaluation
% purpose as we will discuss in Section \ref{SimuResults}, we only
% consider records with race of black and white in our analysis.
%The final datasets for our empirical study has $n=2670$ records.

In these data, $Y$ does not exhibit strong residual
spatial dependence after accounting for other variables. Therefore,
%availability of an outcome variable with strong spatial dependence
for a more thorough test of the analytical validity of the\vspace*{1pt}
synthetic data sets, we also generate a surrogate cause of death
variable, $\tilde{Y}$, that exhibits spatial clustering and is
dependent on several nongeographic variables. To do so, we generate
outcomes as follows:
%We select $S$, $R$ and $A$ from the
%dataset as the variables composing the spatial logistic model and
%simulate binomial outcomes $Y$ using the actual variable values
%according to a spatial logistic regression model. We have
%
%e1 #&#
\begin{equation}\label{eqspatial}
\tilde{Y}_i \sim \operatorname{Bern}(\pi_i),
\end{equation}
where
$\operatorname{logit}(\pi_i) = 0.02 + \mathrm{Sex}_i + \mathrm{Race}_i
+ 0.003 \operatorname{Age}_i +
w(\mathbf{s}_i)$, $\mathbf{s}_i = (\lambda_i, \phi_i)$,\break and~$w(\mathbf{s})$ is a mean
zero Gaussian process with exponential covariance function
$C(\mathbf{s},\mathbf{s}^\prime) = \sigma_e^2 \exp(-\phi_e
\|\mathbf{s}-\mathbf{s}^\prime\|_2)$. We set the parameters of the
exponential
covariance function to $\sigma_e^2=2$ and $\phi_e=0.06$, so
that the effective range (i.e., the distance at which the spatial correlation
drops to 0.05) is about $-\log(0.05)/\phi_e = 50$, which equals
half of the overall range of the latitudes and longitudes in Table
\ref{tabdescription}. The coefficients of the covariates are
specified so that the covariates have strong effects.
%In simulations
%not reported here, we used other values for the coefficients and found
%qualitatively similar results to those observed in
%the simulations with $\tilde{Y}$ and with $Y$.
All results that follow use $\tilde{Y}$ in place of the
actual cause of death $Y$; see the online supplement
[\citet{WangReiter2011Supp}] for
selected results based on $Y$. For both $\tilde{Y}$ and $Y$, the
results are qualitatively similar, in that the actual spatial
relationships (or lack thereof) in
the original data are approximately preserved in the synthetic data sets.
%the simulations with $\tilde{Y}$ and with $Y$.

%We simulate
%$Y$ to ensure availability of an outcome variable with a known and
%strong spatial dependence for use in illustrating the analytical
%validity of the synthetic datasets. cause of death does not exhibit
%strong residual
%spatial dependence after accounting for other variables.

%s3.1 #&#
\subsection{Generation of synthetic data} \label{MortalityGeneration}

%We apply the two types of synthesis models of Section
%and nongeographic quasi-identifiers.

%CART models \citet{breiman1984} are flexible tools for estimating
%the conditional distribution of a univariate outcome given
%multivariate predictors. Essentially, the CART model partitions the
%predictor space so that subsets of units formed by the partitions
%have relatively homogeneous outcomes. The partitions are found by
%recursive binary splits of the predictors. The series of splits can
%be effectively represented by a tree structure, with leaves
%corresponding to the subsets of units. \citet{reitercart}
%describes
%how CART models can be used to generate partially synthetic data.

%create synthetic datasets.
% First specify the order of
%two CART models that generate latitude and longitude, respectively.
%imputation
%using CART models. The basic idea is to impute the variables in
%decreasing order of dependency on each other. For latitude and
%longitude, their mutual dependencies are about the same. In our
%empirical study, the order appears to have little effect on the
%utility and risk measures.

%HAO: MAKE SURE THAT THE PARAGRAPH BELOW IS TRUE, IN THAT WE DID ALL
%OF THE THINGS THAT WE SAID. {\color{red} (Yes!)}

We examined several methods for simulating latitude and longitude,
including mixtures of bivariate regressions, bivariate partition
models [\citet{Death08}] using the ``mvpart'' function in R, Bayesian
additive regression trees [\citet{Chipman10BART}], and classification
and regression trees (CART) [\citet{breiman1984}]. Among these, the
CART synthesizer resulted in data sets with a~desirable profile in
terms of low disclosure risks and high data usefulness. Furthermore,
the CART synthesizer is fastest computationally and easy to
implement, as it requires minimal tuning. It scales to large
data sets with many predictors and many observations.
%The other
%models were less desirable in terms of utility, risk and
%computational cost than the CART models. In particular, note that
%The multivariate response models have no ordering issue of longitude
%and
%latitude;
In comparison to the CART synthesizer, the Bayesian trees and mixture
model synthesizers were far more computationally\vadjust{\goodbreak} demanding, and the
bivariate partition model synthesizer resulted in unacceptably high
disclosure risks. We therefore present
results only for the CART synthesizer, which we now
summarize; see \citet{reitercart} for further information on CART
synthesizers.

%We sequential CART models for response surface modeling,
%alternative approaches include some multivariate response models
%such as bivariate partition models \citet{Death08} and bivariate
%mixtures regression models, and sequences of other type of
%univariate parametric or nonparametric models such as Bayesian CART
%models \citet{Chipman97bayesiancart} and BART models
%multivariate response models have no ordering issue of longitude and
%%latitude; in empirical studies, however, we find the synthetic data
%from multivariate partition models (fitted using ``mvpart'' function
%in R) have higher risks than sequential CART models. The Bayesian
%approaches and mixture models require very demanding computational
%burden in the examples we studied as the number of synthesized data
%is often in thousands or even larger. In contrast, and though with
%ordering issue, CART models are easy to be implemented, and
%scaled-up to large datasets with many predictors and many
%%observations. The empirical studies have verified the resulting
%utility and risk of synthetic datasets.

%Classification and regression trees (CART) have been used to create
%synthetic attributes \citet{reitercart,reiterisr,sampsynthesis}.
%We
%utilize them here to create synthetic locations.
%Without loss of generality,
%suppose the
%order is first longitude, denoted by $\lambda$, and then latitude,
%denoted by $\phi$.
Let $\mathbf{x}$ include all nongeographic attributes in Table
\ref{tabdescription} and $\tilde{Y}$. First, we fit a
regression tree of longitude on $\mathbf{x}$. Label the tree as
$\mathcal{T}_{\lambda}$, where $\lambda$ stands for
longitude. Let $L_{\lambda,w}$ be the $w$th leaf in
$\mathcal{T}_{\lambda}$, and let $\bflambda_{L_{\lambda,w}}$ be the
$n_{L_{\lambda,w}}$ values of $\lambda$ in leaf $L_{\lambda,w}$. In
each $L_{\lambda,w}$, we draw $n_{L_{\lambda,w}}$ values from
$\bflambda_{L_{\lambda,w}}$ using the Bayesian bootstrap
[\citet{rubin1981}]. We then smooth
%These sampled values are the replacement
%imputations for the $n_{L_{\lambda,w}}$ units that belong to
%$L_{\lambda,w}$. Repeating the Bayesian bootstrap in each leaf of
%the longitude tree results in the $l$th set of synthetic longitudes,
%$\tilde{\bflambda}^{(l)}$.
%To avoid releasing values of the original longitudes, in each leaf
the density of the bootstrapped values using a Gaussian kernel
density estimator with bandwidth $h_\lambda$ and support over the
smallest to the largest value of $\bflambda_{L_{\lambda,w}}$. To get
a synthetic longitude for the $i$th unit, we trace down
$\mathcal{T}_{\lambda}$ based on the unit's values of $\mathbf
{x}_i$, and
we sample randomly from the estimated mixture density in that unit's
leaf.
%We repeat the
%process in each leaf of $\mathcal{Y}_{\lambda}$, resulting in the
The result is a set of synthetic longitudes,
$\tilde{\bflambda}{}^{(l)}$.

%To simulate geography, we first grow two CART trees: tree
%$\mathcal{Y}_\lambda$ of $\lambda$ on all non-geographical variables
%listed in Table \ref{tabdescription} and $Y$, and tree
%$\mathcal{Y}_\phi$ of $\phi$ on all nongeographical variables and
%$\lambda$. To limit disclosure risk, we require the smallest leaf
%node size to be five. All CART models are fit in R using the
%``tree'' function. After obtaining $\mathcal{Y}_\lambda$ and
%$\mathcal{Y}_\phi$, we then apply the Bayesian bootstrap and the
%kernel smoothing steps to simulate geography. It takes about 1
%minutes on a standard desktop computer to fit the trees and generate
%five synthetic datasets. To illustrate and compare the utility and
%disclosure risks under different bandwidths, we consider three
%bandwidth levels $h_\lambda=h_\phi=1, 5,\textrm{ and } 10$.

Next, we
% latitude $\phi$ using the similar
%procedure with some minor modifications in order to maintain
%consistency with the $\tilde{\bflambda}^{(l)}$. First, we
fit the regression tree of latitude on $\mathbf{x}$ and the true
$\lambda$; label the tree as~$\mathcal{T}_\phi$, where $\phi$ stands
for latitude. To locate the $i$th
person's leaf in~$\mathcal{T}_\phi$, we use
$\tilde{\lambda}^{(l)}_i$ in place of $\lambda_i$. For units with
combinations of $(\mathbf{x}_i,\tilde{\lambda}^{(l)}_i)$ that do not belong
to one of the leaves of $\mathcal{T}_\phi$, we search up the tree
until we find a node that contains the combination, and treat that
node as if it were the unit's leaf. Once each unit's leaf is
located, values of $\phi^{(l)}_i$ are generated using the Bayesian
bootstrap and kernel density procedure with bandwidth~$h_\phi$. The
result is a set of synthetic latitudes, $\tilde{\bfphi}{}^{(l)}$, and,
therefore, synthetic locations $\tilde{\mathbf{s}}{}^{(l)} =
(\tilde{\bflambda}{}^{(l)}, \tilde{\bfphi}{}^{(l)})$.

We repeat the process of generating $\tilde{\mathbf{s}}{}^{(l)}$
%$(\tilde{\bflambda}^{(l)}, \tilde{\bfphi}^{(l)})$
independently $m$ times, resulting in the collection of partially
synthetic data sets, $D^{(l)} = \{\mathbf{x}, \tilde{\mathbf{s}}^{(l)}
\}$ where $l=1,\allowbreak\ldots,m$. With no further synthesis of $\mathbf{x}$, these
$m$ data sets would be released to the public.

We also performed the synthesis by generating latitude first and longitude
second. As reported in the online supplement
[\citet{WangReiter2011Supp}], this ordering results in
slightly decreased disclosure risks and slightly worse data
utility. We recommend that data
stewards try both orderings and choose the one
that results in the more desirable risk-utility profile. For general
discussions on the order of synthesis, see
\citet{reitercart} and \citet{reitcaiola}.

%In general, data stewards can improve protection by simulating both
%geography and certain components of $\mathbf{x}$ deemed readily
%available to
%ill-intentioned users; these are called quasi-identifiers. Typical
%examples of quasi-identifiers include age, race, sex, marital status,
%and housing characteristics.
%Typically, quasi-identifiers do not
%constitute all attributes in
%$\bx$.
%For example, ill-intentioned users may not know study
%%participants!`\={} answers to questions about opinions or highly
%personal facts.
%To differentiate types of variables, we let $\bx= (\bx_{1},
%$\bx_{2}$ are non-identifiers.

%We fit CART models for $\bx_{1}$ as a function of
%$\bx_{2}$ and $(\phi,\lambda)$.
%This provides the following synthetic data generating
%process. First, simulate new longitudes and latitudes
%$(\tilde{\lambda},\tilde{\phi})$ using the methods in Section
%proper distribution when next simulating the
%quasi-identifiers. Second, simulate new values of quasi-identifiers
%%$\bx_{i,1}$ using the methods in \citet{reitercart}.
%Each $\bx_{i,1}$
%is simulated based on its
%$(\tilde{\lambda},\tilde{\phi})$ from Step 1.

We also investigate simulating both geography and nongeographic
identifiers to further improve confidentiality protection.
Specifically, we simulate values of race ($R$) and age ($A$) in
addition to $(\lambda,\phi)$. We choose these two variables because
(i) in many
applications, age and race might be considered available to
ill-intentioned users and hence prominent candidates for
disclosure protection, (ii) their distributions clearly depend on
location in the NC mortality data, and (iii) they encompass the generic
modeling challenges of a continuous and a categorical
variable.

The process proceeds as follows. Simulate $(\tilde{\lambda},\tilde
{\phi})$ using the CART
synthesizers as before, but excluding $R$ and $A$ from $\mathbf{x}$. We
simulate new values of~$A$ using a CART synthesizer fit with
$(\mathbf{x},\lambda,\phi)$. Each $A_i$ is simulated based on its
$(\tilde{\lambda}_i,\tilde{\phi}_i)$. We simulate new values of $R$
using a\vspace*{1pt} CART synthesizer fit with~$(\mathbf{x},\allowbreak\lambda,\phi, A)$. Each
$R_i$ is simulated based on its $(\tilde{\lambda}_i,\tilde{\phi}_i,
\tilde{A}_i)$.

%All
%trees are pruned so that there are at least 5 observed values in
%each leaf. As before, we consider $h_\lambda=h_\phi= 1, 5,\textrm{
%and } 10$ for generating $(\tilde{\lambda}_i,\tilde{\phi}_i)$. In
%addition, we set the bandwidth for generating $\tilde{A}_i$ equal
%$2$. In all simulations, the number of synthetic datasets $m$
%equals 5.

For all trees, we require the smallest node size to be at least
five, and we cease splitting a leaf when the deviance of values in
the leaf is less than 0.0001; see Section \ref{discussion} for discussion of selecting
these tuning parameters. All CART models are fit in R using the
``tree'' function. The bandwidth sizes are directly related to the
analytical utility and disclosure risks of the synthetic data sets.
Here, we investigate the risk-utility trade-offs for three
bandwidths: $h_\lambda=h_\phi\in\{1, 5, 10\}$. We set the
bandwidth for generating $\tilde{A}$ equal to $2$. We generate $m= 5$
synthetic data sets.

%We seek to simulate these variables with There are a number of
%potential
%approaches to simulate quasi-identifiers such as Gaussian process
%model with a spatial covariance matrix; we discuss this in Section

%s3.2 #&#
\subsection{Evaluation of confidentiality protection}

\begin{figure}

\includegraphics{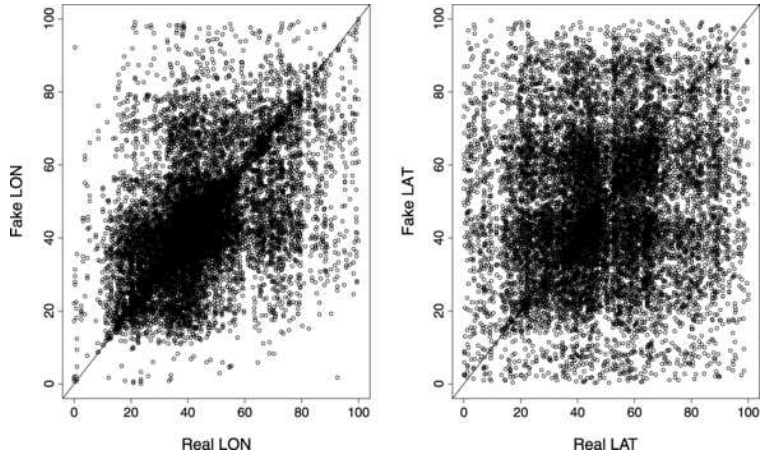}

\caption{Scatter plots of synthetic longitudes (left) and latitudes
(right) against real ones
under the synthesis model that imputes geography only with $h=1$. }
\label{fig2}
\vspace*{-3pt}
\end{figure}

For an initial evaluation of the protection
engendered by simulation, we plot $(\tilde{\lambda},\tilde{\phi})$
against $(\lambda,\phi)$ for one simulated data set when only
geography is imputed with $h=1$; see Figure \ref{fig2}. Clearly,
$(\tilde{\lambda}_i,\tilde{\phi}_i)$ can vary greatly from
$(\lambda_i,\phi_i)$. However, Figure \ref{fig2} is a crude
evaluation, as intruders can utilize information from the multiple
synthetic data sets and possibly other information to attempt
disclosures.

%
%f1 #&#

We now outline frameworks for evaluating disclosure
risks. We begin with an approach for quantifying how much intruders
can learn about actual geographies from the synthetic data.

%s3.2.1 #&#
\subsubsection{Risk of geography disclosure} \label{SectionRiskAttri}

%This section proposes a novel framework for quantifying the
%risks that intruders can learn actual geographies from synthetic data.
In this section we assume that geography is the only synthesized
variable, although the general ideas and~ap\-proach apply to other
attributes and with additional synthetic data.
%For a collection of $n$ sampled units, let $\bx_i$ be the $i$th
%person's vector of non-geographic attributes, such as their age,
%gender and race.

%The data steward releases $l=1,\cdots,m$ synthetic datasets including
%all
%$n$ sampled units. %Let $\bz_i^{(l)}$ be the released values for
%person $i$ in partially synthetic datasets $l$.
%We assume that the data steward synthesizes only values of latitudes
%and
%longitudes.

Let $\tilde{\mathbf{s}}_i =
(\tilde{\mathbf{s}}_i^{(1)},\ldots,\tilde{\mathbf{s}}_i^{(m)})$;
let $\tilde{\bS}$
be the collection of $\tilde{\mathbf{s}}_i$ for all $n$ persons in the
sample; and,
% let $\bX$ be the collection of $\bx_i$ for
%all $n$ persons in the sample. Finally,
let $\bS$ include\vadjust{\goodbreak} all $n$ original values of $\mathbf{s}_i$.
%Let $\bs_i =(\phi_i,\lambda_i)$ and $\tilde{\bs}_i^{(l)}
%=(\tilde{\phi}_i^{(l)},\tilde{\lambda}_i^{(l)})$ denote the observed
%and $l$th set of synthetic locations, respectively, of person $i$.
Let $\bS_{-i}$ be all the original geography except for that of the
$i$th person. Let $M$ represent any meta-data released by the data
steward about the synthesis models, for example, the code for the
computer program that generated that synthetic data (without the
original data or parameter estimates). Let $I$
represent the intruder's prior information on persons' geography in
the sample, for example, $I$ might include $\bS_{-i}$. Either $M$
or $I$ could be empty.

We posit an intruder whose goal is to estimate $\mathbf{s}_i$ for one or
more target records in the database.
%The data steward generating synthetic geography might release meta-data
%about the synthesis process to help analysts determine if their
%analyses are reasonably supported in the synthetic data. Let $M$
%represent the meta-data released about the models used to generate
%the synthetic data. The $M$ could include, for example, the code for
%the models used to generate the synthetic data. $M$ could also be
%empty.
%
%The intruder might have prior information, $I$, on persons'
%geography in the sample. For example, the intruder may know persons'
%block address. $I$ could be empty.
Specifically, for any record $i$, the intruder seeks the posterior
distribution of $\mathbf{s}_i$ given $(\bX, \tilde{\bS}, M, I)$.
%$P( \bs_i \mid\bX, \tilde{\bS}, M,I )$.
With this posterior distribution, the intruder could identify high
density regions for the unknown $\mathbf{s}_i$, which, if precise enough,
could be used to pinpoint the true location of the target
individual. Using Bayes' rule, we have
%
%e2 #&#
\begin{equation} \label{eqrisk}
P( \mathbf{s}_i |\bX, \tilde{\bS}, M, I) \propto P( \tilde
{\bS} |
\bX,\mathbf{s}_i, M, I) P( \mathbf{s}_i |\bX, M, I),
\end{equation}
where $ P( \mathbf{s}_i |\bX, M, I)$ represents the intruder's prior
beliefs about $\mathbf{s}_i$.

The information in $M$ and $I$ play central roles in the likelihood
function~$P( \tilde{\bS} |\bX,\mathbf{s}_i, M, I )$. For example,
suppose that $M$ contains the code of the computer program used to
generate the synthetic data (without original data or parameter
estimates). If $I$ includes $\bS_{-i}$, the intruder could take
guesses at $\mathbf{s}_i$
according to his or her prior distribution and, with the resulting
guess of $\bS$, determine the likelihood of $\tilde{\bS}$. If
instead $I$ contains only a portion of the geographies or is empty,
as are likely to be the cases in practice, the computation of the
likelihood becomes much more complex and uncertain, since the
intruder needs to guess at multiple unknown geographies. In such
cases, one simple approximation of the distribution for $\mathbf{s}_i$ is
the convex hull of the set $\tilde{\mathbf{s}}_i$. Given the
variation in
Figure \ref{fig2}, these regions in the mortality data could
be quite large.

The intruder's prior distribution is also a key determinant of the
posterior distribution of $\mathbf{s}_i$. An intruder may know the
locations of all individuals in the population with certain
characteristics contained in $\mathbf{x}_i$, and the prior distribution
could be uniform over those locations. An intruder who knows~$\mathbf{x}$
and $\bS_{-i}$ could estimate a model from these data to predict
$\mathbf{s}_i$,
and use that as a prior distribution. An intruder with no external
information might use a uniform distribution on the map of possible
locations. Unfortunately, it is nearly impossible for the data
steward to know the information possessed by the intruder. Hence, it
is prudent for the data steward to consider disclosure risks under a
variety of assumptions about the intruders' knowledge---including
very extensive prior knowledge, which represents possible worst case
scenarios---as
we now demonstrate.
%Another alternative is to use a bivariate
%normal distribution centered at the average of $\tilde{\bs}_i^{(l)}$

% SOMETHING HERE HIPAA stuff.

%in the CART synthesizer,
%suppose that the data steward releases the exact specification of the
%CART
%models, including the splitting rule of each node and the kernel
%bandwidth and the intruder also
%knows all the geography of the $n$ persons in the sample except for
%the targeted $i$th person, i.e. $M$ has complete information and
%$I$ is equal to $\bS_{-i}$. In such case, the intruder may use
%sampling techniques to estimate $\bs_i$ as we will illustrate in the
%empirical study.
% If the
%intruder knows nothing about the synthetic data generating
%process nor the prior information about geography, i.e., both $M$
%and $I$ are empty, the intruder is almost
%
%s possible to assume $P( \tilde{\bS} \mid\bX
%,\bs_i, M,I ) \propto N(\tilde{\bs}_i; \bs_i,\sigma^2)$, and the
%average of $\tilde{\bs}_i$ to estimate $\bs_i$. We also consider a
%very risky scenarios about $M$ and $I$, that is the data steward
%releases
%the exact specification of the CART models, including the splitting
%rule of each node and the kernel bandwidth and the intruder also
%knows all the geography of the $n$ persons in the sample except for
%the targeted $i$th person, i.e. $M$ has complete information and
%$I$ is equal to $\bS_{-i}$. In such case, the intruder may use
%sampling techniques to estimate $\bs_i$ as we will illustrate in the
%empirical study.

Using the CART synthesizer, we consider two scenarios for the NC
mortality data: a high-risk scenario in which the intruders know
everything except for one target's $\mathbf{s}_i$, that is, $\bX$ and
$\bS_{-i}$, and a low-risk scenario in which the intruder does not
know any records' geographies. We\vadjust{\goodbreak} assume that $M$ includes
everything about the trees except the individual geographies in the
nodes, that is, the data steward releases the splitting rules for
each tree and the kernel bandwidths.
%This is arguably more
%information than many data stewards would comfortably release;
%however, it is prudent for data stewards to act as if intruders have
%this knowledge, particularly if it is plausible that they know
%$\bS_{-i}$.
% comit facilitates estimation of posterior distributions of each $
%the dataset.
For the risky scenario, we assume the intruder's prior distribution
is uniform on a grid over a small area containing the target's true latitude
and longitude, and estimate equation (\ref{eqrisk}) using
importance sampling; see the online supplement
[\citet{WangReiter2011Supp}] for details.
Because the small area contains the true value, this prior distribution
represents strong intruder prior knowledge.
%%grid of values is carefully chosen to be
%informative based on the information in $\bX$, $M$ and $I$; see the
%online supplement \citet{WangReiter2011Supp} for details. There
%are
We note that other specifications for the prior distribution could
change the value of the risk measure.

%
%t2 #&#
\begin{table}[b]
\caption{Summary of geography disclosure risks for the low
risk and high risk scenarios for different
bandwidths $h$ when synthesizing only geography. For each risk
measure, $\alpha_{0}$ is the minimum, $\alpha_{25}$ is the
first quartile, and $\alpha_{50}$ is the median}
\label{tabAttrRiskGeoOnly}
\begin{tabular*}{\tablewidth}{@{\extracolsep{\fill}}lcd{2.1}d{3.1}d{3.1}d{2.1}d{3.1}
d{3.1}d{2.1}d{3.1}d{3.1}@{}}
\hline
& & \multicolumn{3}{c}{$\bolds{h = 1}$} & \multicolumn{3}{c}{$\bolds{h=5}$} &
\multicolumn{3}{c}{$\bolds{h=10}$}\\[-4pt]
& & \multicolumn{3}{c}{\hrulefill} & \multicolumn{3}{c}{\hrulefill} &
\multicolumn{3}{c@{}}{\hrulefill}\\
\textbf{Scenario} & \textbf{Risk} & \multicolumn{1}{c}{$\bolds{\alpha_{0}}$}
& \multicolumn{1}{c}{$\bolds{\alpha_{25}}$}
& \multicolumn{1}{c}{$\bolds{\alpha_{50}}$} & \multicolumn{1}{c}{$\bolds{\alpha_{0}}$}
& \multicolumn{1}{c}{$\bolds{\alpha_{25}}$}
& \multicolumn{1}{c}{$\bolds{\alpha_{50}}$} & \multicolumn{1}{c}{$\bolds{\alpha_{0}}$} &
\multicolumn{1}{c}{$\bolds{\alpha_{25}}$} & \multicolumn{1}{c@{}}{$\bolds{\alpha_{50}}$}\\
\hline
Low & ${R}_{1}$ & 4.2 & 15.6 & 21.6 & 3.6 &
15.4 & 20.8 & 3.8 & 16.0 & 21.4\\
& ${R}_{2}$ & 36 & 384 & 680 & 27 & 373 & 640 & 43 & 393 &
674\\
[4pt]
High & ${R}_{1}$ & 0.0 & 4.3 & 9.7 & 2.1 &
9.7 & 14.2 & 3.4 & 13.1 & 17.8\\
& ${R}_{2}$ & 0 & 34 & 159 & 4 & 149 & 327 & 13 & 253 &
474\\
\hline
\end{tabular*}
\end{table}

To summarize how much the CART synthesis protects geographies, we
create two risk metrics. Let $(\phi_{i,t},\lambda_{i,t})$ be a draw
from $P( \mathbf{s}_i |\bX, \tilde{\bS}, M, I)$. The metrics are
\begin{eqnarray*}
{R}_{1} &=& \biggl[\int\{ (\phi_i-\phi_{i,t})^2 +
(\lambda_i-\lambda_{i,t})^2 \} P( \phi_{i,t},\lambda_{i,t} |\bX,
\tilde{\bS}, M,I ) \,\ud\phi_{i,t} \,\ud\lambda_{i,t}\biggr]^{1/2},\\
{R}_{2} &=& \mbox{number of actual cases
in circle centered at }
(\phi_{i},\lambda_{i}) \quad\mbox{with radius } \uR_1.
\end{eqnarray*}
%
%where $(\phi_{i,t},\lambda_{i,t})$ is the observed latitude and
%longitude of unit $i$.
Here, ${R}_1$ measures the average Euclidean distance between the
intruder's guess of geography and the actual geography. Larger values
of ${R}_1$ (up to a~max of $100\sqrt{2}$)
indicate larger uncertainty in predicting $\mathbf{s}_i$, so that intruders'
predictions are more likely to be further away from the true
geography; thus, larger values of ${R}_1$ indicate smaller
disclosure risks. Larger values of~${R}_{2}$
indicate that many actual locations (up to a max of $n=2\mbox{,}670$) are
reasonable guesses at
$\mathbf{s}_i$, thus smaller disclosure risks.

Table \ref{tabAttrRiskGeoOnly} displays summary statistics for
${R}_{1}$ and ${R}_{2}$ for all $n=2\mbox{,}670$ records in
the database. For the low-risk scenario, the medians of
${R}_1$ for all three bandwidth values are around 21 distance
units, and the medians of ${R}_2$ are around 670 units,
indicating that most $\mathbf{s}_i$ are estimated with sizable uncertainty.
In this scenario, each person's ${R}_{1}$-radius circle
contains at least 27 other cases. Interestingly, for this scenario,
increasing the bandwidth does not substantially increase the
uncertainty in $\mathbf{s}_i$. For the high-risk scenario, the
intruder can
estimate $\mathbf{s}_i$ with better accuracy than in the low-risk scenario.
Here, both ${R}_{1}$ and ${R}_{2}$ decrease with $h$.
In fact, when $h=1$, there are individuals in the data who are alone
in their ${R}_{1}$-radius circles. The boxplot of Figure 2 in
the online supplement [\citet{WangReiter2011Supp}] provides additional
information about the distributions of ${R}_{1}$ and
${R}_{2}$, including those under different scenarios when
generating latitude first and longitude second.

%This
%is because the multi-modal likelihood function of $\bs_i$
%becomes flatter as $h$ increases.

%s3.2.2 #&#
\subsubsection{Risk of identification} \label{SectionRiskID}
The approach in Section \ref{SectionRiskAttri} can be used to
estimate posterior distributions of any attribute, of which location
is one example. Often, however, data stewards want to assess the
risks that individuals in the released data can be re-identified. To
quantify these risks, we now compute probabilities of identification
[\citet{dl89}, \citet{fms97}, \citet{reiter05}] by
adapting the approach of
\citet{drechreiter08} and \citet{reitermitra07} for synthetic
geographies.
% The description
%below is paraphrased from \citet{reitermitra07}.

In this approach, the data steward mimics the behavior of an
intruder who possesses the true values of the quasi-identifiers,
including geographies, for selected target records (or even the
entire database). To illustrate, suppose the intruder has a vector
of information, $\mathbf{t}$, on a particular target unit in the population
which may or may not correspond to a unit in the $m$ released
synthetic data sets, $D = \{D^{(1)}, \ldots,D^{(m)} \}$. Let $t_0$ be
the unique identifier (e.g., the individual's name) of the target,
and let $d_{j0}$ be the (not released) unique identifier for record
$j$ in $D$, where $j = 1,\ldots, n$. The intruder's goal is to
match unit $j$ in $D$ to the target when $d_{j0} = t_0$, and not to
match when $d_{j0} \neq t_0$ for any $j \in D$.

Let $J$ be a random variable that equals $j$ when $d_{j0} = t_0$ for
$j \in D$ and equals $n + 1$ when $d_{j0} = t_0$ for some $j \notin
D$. The intruder thus seeks to calculate the $P(J = j|
\mathbf{t},D,M)$ for $j = 1,\ldots,n + 1$. He or she then would decide
whether or not any of the identification probabilities for $j = 1,
\ldots, n$ are large enough to declare an identification. Let $\bT$
be all original values of the variables that were synthesized.
Because the intruder does not know the actual values in $\bT$,
he or she should integrate over its possible values when computing
the match probabilities. Hence, for each record in $D$ we compute
\[
P(J = j|\mathbf{t},D,M) = \int \operatorname{Pr}(J =
j|\mathbf{t},D,\bT,M, I)\operatorname{Pr}(\bT|\mathbf{t},D,M, I)\,dT.
\]
This integral can be approximated using Monte Carlo
approaches; details are in the online supplement. Once again, the
data steward must make assumptions about $I$, the information the
intruder knows about the targets.

%construction suggests a Monte Carlo approach to estimating each
%$P(J = j|\bt,D,M, I)$. First, sample a value of $\bT$ from
%$P(\bT|\bt,D,M, I)$. Let $\bT_{new}$ represent one set of simulated
%values. Second, compute $P(J = j|\bt,D,\bT= \bT_{new},M, I)$ using
%exact or, for continuous synthesized variables, distance-based
%matching assuming $\bT_{new}$ are collected values. This two-step
%process is iterated R times, where ideally R is large, and equation
%(\ref{eqIDrisk}) is estimated as the average of the resultant $R$
%values of $P(J = j|\bt,D,\bT= \bT_{new},M, I)$. When $M$ has no
%information, the malicious user can treat the simulated values as
%plausible draws of $\bT$.

Data stewards can summarize the risks for the entire data set using
functions of these match probabilities [\citet{reiter05}]. Let $c_j$
be the number of records in the data\vadjust{\goodbreak} set with the highest match
probability for the target $\mathbf{t}_j$. Let $g_j=1$ if the true match
is among the $c_j$ units, and $g_j = 0$ otherwise. The expected
match risk equals $\sum_j(1/c_j)g_j/n$. The true match risk equals
$\sum_jk_j/n$, where $k_j = 1$ when $c_j
g_j = 1$, and $k_j = 0$ otherwise. The false match risk equals
$\sum_j f_j (1 - g_j)/ \sum_j f_j $, where $f_j = 1$ when $c_j=1$
and $f_j = 0$ otherwise. Effective disclosure limitation techniques
have low expected and true match risks, and high false match risks.

Using the mortality data, we consider three scenarios with different
information in $M$. In the first $M$ contains everything, that is,
details of the CART models, the splitting rules and the real data
values in each leaf and internal node. Essentially, $M$ is a data
simulator that enables analysts to generate new synthetic data sets
using the same process as the data steward. In the second $M$
contains descriptions of the CART models, but not the specific
splitting rules nor the real data values in each leaf and internal
node. Essentially, this is akin to releasing the code used to
simulate data without providing any parameter values for it. In the
third $M$ is empty, that is, the data steward says nothing about how
the data were collected.

For all scenarios, we suppose that intruders have a file containing
the true values of sex, race, marital status, age and geography for
all $n=2\mbox{,}670$ units in the data set, and that they seek to match
records in $D$ to this file. We also suppose that the intruder knows
which records were in the sample, so that $P(J = 2671|\mathbf{t},D,M) =0$.
%Thus, or any $\bt$ in the sample, we compute
%$P(J = j |\bt,D,M)$ for $j=1,\cdots, n$.
We compute each target's probability independently of other targets'
probabilities and match with replacement.

%%\begin{table}[tbp]
%%\center
%%\caption{Summary of risk measures under different scenarios
%%when synthesizing only geography and when
%%synthesizing geography, age, and race. Here, $E$ is expected
%%match risk, $T$ is true match risk, and $F$ is false match
%%risk. Results based on one simulation run per scenario.}
%%\label{tabIdRiskGeoNongeo}
%%\footnotesize
%%
%%\begin{tabular}{lrrrrrrrrrrr}
%%\hline
%% & \multicolumn{3}{c}{$h=1$} & & \multicolumn{3}{c}{$h=5$} & &
%%Information in $M$ & $E$ & $T$ & $F$ & & $E$ & $T$ & $F$ & & $E$ &
%$T$ & $F$ \\
%%\hline
%%\multicolumn{3}{l}{\underline{Synthesizing geography only}} \\
%%$ $Empty & 568 & 396 & .76& & 494 & 326 & .78 & & 459 & 288
%& .80\\
%%\\
%%$ $Code, no parameters & 573 & 504 & .78 & & 508 & 432 &
%.80& & 499 & 412 & .82\\
%%\\
%%% & \multicolumn{3}{c}{All $h$} & & \multicolumn{3}{c}{} & &
%%$ $Everything & 896 & 845 & .48 & & - & - & - & & - & - & -\\
%%
%%\\
%%\multicolumn{3}{l}{\underline{Synthesizing geography, age, and race}}
%%$ $Empty & 27 & 21 & .98 & & 23 & 19 & .99& & 20 & 16& .99\\
%%\\
%%$ $Code, no parameters & 25 & 25 & .99 & & 21 & 21 & .99 & &
%13 & 13 & .99\\
%%\\
%%% & \multicolumn{3}{c}{All $h$} & & \multicolumn{3}{c}{} & &
%%$ $Everything & 91 & 90 & .94 & & - & - & - & & - & - & -\\
%%\hline
%%\end{tabular}
%%\end{table}

%
%t3 #&#
\begin{table}[b]
\tabcolsep=4pt
\caption{Summary of risk measures under different scenarios when
synthesizing only geography and when synthesizing geography, age
and race. Here, $E$ is expected match risk, $T$ is true match risk,
and $F$~is false match risk. Results based on one simulation run
per scenario}
\label{tabIdRiskGeoNongeo}
\begin{tabular*}{\tablewidth}{@{\extracolsep{\fill}}ld{1.3}d{1.3}cd{1.3}d{1.3}cd{1.3}d{1.3}c@{}}
\hline
& \multicolumn{3}{c}{$\bolds{h=1}$} & \multicolumn{3}{c}{$\bolds{h=5}$} &
\multicolumn{3}{c}{$\bolds{h=10}$}\\[-4pt]
& \multicolumn{3}{c}{\hrulefill} & \multicolumn{3}{c}{\hrulefill} &
\multicolumn{3}{c@{}}{\hrulefill}\\
\textbf{Information in} $\bolds{M}$ & \multicolumn{1}{c}{$\bolds{E}$}
& \multicolumn{1}{c}{$\bolds{T}$} & \multicolumn{1}{c}{$\bolds{F}$}
& \multicolumn{1}{c}{$\bolds{E}$} & \multicolumn{1}{c}{$\bolds{T}$}
& \multicolumn{1}{c}{$\bolds{F}$} & \multicolumn{1}{c}{$\bolds{E}$}
& \multicolumn{1}{c}{$\bolds{T}$} & \multicolumn{1}{c@{}}{$\bolds{F}$} \\
\hline
\multicolumn{10}{@{}c@{}}{\textit{Synthesizing geography only}} \\[4pt]
Empty & 0.21 & 0.15 & 0.76& 0.19 & 0.12 & 0.78 & 0.18 & 0.11 &
0.80\\
Code, no parameters & 0.21 & 0.19 & 0.78 & 0.19 & 0.16 & 0.80& 0.18 & 0.15 & 0.82\\
% & \multicolumn{3}{c}{All $h$} & \multicolumn{3}{c}{} & &
Everything & 0.34 & 0.32 & 0.48 & \multicolumn{1}{c}{--} & \multicolumn{1}{c}{--} & \multicolumn{1}{c}{--} & \multicolumn{1}{c}{--}
& \multicolumn{1}{c}{--} & \multicolumn{1}{c@{}}{--}\\[6pt]
\multicolumn{10}{@{}c@{}}{\textit{Synthesizing geography, age and race}}
\\[4pt]
Empty & 0.010 & 0.008 & 0.98 & 0.008 & 0.007 & 0.99& 0.007 &
0.006& 0.99\\
Code, no parameters & 0.009 & 0.009 & 0.99 & 0.008 & 0.008 &
0.99 & 0.005 & 0.005 & 0.99\\
% & \multicolumn{3}{c}{All $h$} & \multicolumn{3}{c}{} & &
Everything & 0.034 & 0.034 & 0.94 & \multicolumn{1}{c}{--} & \multicolumn{1}{c}{--} & \multicolumn{1}{c}{--}
& \multicolumn{1}{c}{--} & \multicolumn{1}{c}{--} & \multicolumn{1}{c@{}}{--}\\
\hline
\end{tabular*}
\end{table}

Table \ref{tabIdRiskGeoNongeo} summarizes risk measures in one
set of $m=5$ synthetic data sets for each bandwidth and scenario.
Three general trends are evident; these persist in two additional
runs of the simulation as well. First, the synthesis of age and
race dramatically decreases disclosure risks. Indeed, we suspect
that many data stewards would consider the numbers of true matches
unacceptably high for synthesizing geography only and perhaps
acceptable for synthesizing geography, age and race. Second,
releasing additional information in $M$ increases the disclosure
risks. This trend is particularly pronounced when synthesizing only
geography, and less so when synthesizing geography, age and race.
For the latter synthesis strategy, the incremental risk of releasing
the synthesis code without parameters over releasing nothing is modest,
suggesting that it is worth releasing $M$ to improve
analysts' understanding of the disclosure limitation applied to the
data. Third, the risks tend to increase as the bandwidth for
geography synthesis decreases. This is because larger $h$ implies
larger variances in the synthetic locations.

%s3.3 #&#
\subsection{Evaluation of analytical validity} \label{SimuResults}

As with disclosure risks, the extent to which synthetic data sets can
support analytically valid inferences depends on the properties of
the synthesizer. In this section we examine the quality of
synthetic data inferences for several estimands in the NC mortality
data set. Based on the huge reductions in disclosure risks, we only
consider scenarios with $(\lambda, \phi, R, A)$ synthesized. The
online supplement [\citet{WangReiter2011Supp}] provides
corresponding results with only
$(\lambda, \phi)$ synthesized.

%This section examines the properties of inferences based on the
%imputations that are described in Section \ref{MortalityGeneration}.
%We now investigate the validity of inference using synthetic
%geography, we
%focus on estimands related to geography.
%Note that the properties of
%these inferences are specific to this empirical study; however,
%these results do inform us about the potential benefits and
%limitations of releasing synthetic geography in other datasets with
%geography.

%First, we perform a repeated sampling simulation to evaluate
%the properties of inferences for several descriptive estimands that
%involve geography. For each combination of $h$ and synthesis models, we
%simulate $D$ independently 100 times, each of which comprises $m=5$
%synthetic datasets. All synthetic datasets are generated using on
%methods in Section \ref{MortalityGeneration}.

%
%t4 #&#
\begin{table}
\caption{Summary of simulation results for
descriptive estimands when imputing both
geography and nongeography. $Q$ stands for the
population values; ME and MSE stand for the median and mean
square error of $\bar{q}_5$ across the 100 simulations}
\label{tabdescpt}
\begin{tabular*}{\tablewidth}{@{\extracolsep{\fill}}lcd{2.1}d{2.1}cd{2.1}cd{2.1}c@{}}
\hline
& & &\multicolumn{2}{c}{$\bolds{h=1}$} & \multicolumn{2}{c}{$\bolds{h=5}$} &
\multicolumn{2}{c@{}}{$\bolds{h=10}$}\\[-4pt]
& & &\multicolumn{2}{c}{\hrulefill} & \multicolumn{2}{c}{\hrulefill} &
\multicolumn{2}{c@{}}{\hrulefill}\\
\textbf{Estimand} & \textbf{ZIP} & \multicolumn{1}{c}{$\bolds{Q}$}
& \multicolumn{1}{c}{\textbf{ME}} & \multicolumn{1}{c}{\textbf{MSE}} &
\multicolumn{1}{c}{\textbf{ME}} & \multicolumn{1}{c}{\textbf{MSE}}
& \multicolumn{1}{c}{\textbf{ME}} & \multicolumn{1}{c@{}}{\textbf{MSE}}
\\
\hline
$\%$ black & Z1 & 61.1 & 59.6 & 2.4 & 56.2 & 4.9
& 55.7 & 5.6\\
& Z2 & 41.1 & 43.4 & 1.6 & 41.8 & 1.7 & 40.8 & 1.0\\
& Z3 & 32.6 & 33.7 & 1.7 & 34.0 & 2.0 & 35.5 & 3.0\\
& Z4 & 13.5 & 13.4 & 0.9 & 13.6 & 0.9 & 13.7 & 0.9\\
& Z5 & 46.2 & 44.8 & 1.9 & 43.8 & 2.7 & 42.8 & 3.6\\
& Z6 & 12.9 & 15.2 & 2.5 & 16.4 & 3.7 & 16.3 & 3.5\\
& Z7 & 51.3 & 52.6 & 1.9 & 53.9 & 2.9 & 55.5 & 4.8\\
[6pt]
$\%$ with & Z1 & 19.9 & 18.9 & 1.8 & 20.0 & 1.3 & 21.8 & 2.2\\
\quad$\mbox{educ.}>14.5$ & Z2 & 9.6 & 7.5 & 2.3 & 7.4 & 2.3 & 8.1 & 1.6\\
& Z3 & 10.9 & 9.9 & 1.4 & 10.6 & 0.9 & 13.0 & 2.2\\
& Z4 & 28.3 & 29.3 & 1.6 & 27.8 & 1.1 & 28.1 & 0.8\\
& Z5 & 36.6 & 37.5 & 1.6 & 38.9 & 2.8 & 38.2 & 1.9\\
& Z6 & 25.8 & 26.5 & 1.6 & 26.7 & 1.8 & 27.3 & 2.0\\
& Z7 & 30.9 & 32.8 & 2.6 & 32.0 & 1.9 & 31.8 & 1.7\\
[6pt]
Avg. age & Z1 & 65.8 & 67.6 & 1.9 & 68.7 & 2.9 & 69.1 & 3.4\\
& Z2 & 66.2 & 68.4 & 2.3 & 68.7 & 2.5 & 68.6 & 2.4\\
& Z3 & 71.4 & 70.9 & 0.6 & 70.5 & 1.0 & 70.1 & 1.3\\
& Z4 & 72.5 & 71.6 & 0.9 & 71.3 & 1.2 & 71.2 & 1.3\\
& Z5 & 71.1 & 70.0 & 1.2 & 69.8 & 1.3 & 69.8 & 1.3\\
& Z6 & 72.1 & 71.3 & 1.0 & 71.4 & 0.8 & 71.5 & 0.7\\
& Z7 & 69.4 & 69.4 & 0.5 & 69.5 & 0.5 & 69.7 & 0.6\\
\hline
\end{tabular*}
%
% & & & & \multicolumn{2}{c}{$h=1$} & & \multicolumn{2}{c}{$h=5$} & &
%Estimand & ZIP & $Q$ & & $ME$ & $PD$ & & $ME$ & $PD$ & & $ME$ & $PD$ \\
%56.24 & 0.08 & & 55.69 & 0.09\\
%
% & 27703 & 41.05 & & 43.41 & 0.06 & & 41.85 & 0.02 & & 40.76 & 0.01\\
%
% & 27704 & 32.59 & & 33.72 & 0.03 & & 34.03 & 0.04 & & 35.50 & 0.09\\
%
% & 27705 & 13.47 & & 13.44 & 0.00 & & 13.64 & 0.01 & & 13.74 & 0.02\\
%
% & 27707 & 46.24 & & 44.84 & 0.03 & & 43.81 & 0.05 & & 42.82 & 0.07\\
%
% & 27712 & 12.91 & & 15.23 & 0.18 & & 16.38 & 0.27 & & 16.32 & 0.26\\
%
% & 27713 & 51.32 & & 52.60 & 0.03 & & 53.86 & 0.05 & & 55.50 & 0.08\\
% & 27701 & 19.91 & & 18.90 & 0.05 & & 20.02 & 0.01 & & 21.83 & 0.10\\
%
% & 27703 & 9.61 & & 7.52 & 0.22 & & 7.37 & 0.23 & & 8.14 & 0.15\\
%
%$\%$ of people & 27704 & 10.94 & & 9.92 & 0.09 & & 10.65 & 0.03 & &
%12.97 & 0.19\\
%
% with Educ.$>14.5$ & 27705 & 28.31 & & 29.28 & 0.03 & & 27.78 & 0.02 &
%& 28.15 & 0.01\\
%
% & 27707 & 36.56 & & 37.52 & 0.03 & & 38.88 & 0.06 & & 38.19 & 0.04\\
%
% & 27712 & 25.83 & & 26.55 & 0.03 & & 26.67 & 0.03 & & 27.35 & 0.06\\
%
% & 27713 & 30.92 & & 32.81 & 0.06 & & 32.00 & 0.03 & & 31.79 & 0.03\\
%0.04 & & 69.13 & 0.05\\
%
% & 27703 & 66.24 & & 68.44 & 0.03 & & 68.75 & 0.04 & & 68.64 & 0.04\\
%
% & 27704 & 71.42 & & 70.94 & 0.01 & & 70.49 & 0.01 & & 70.14 & 0.02\\
%
% & 27705 & 72.46 & & 71.62 & 0.01 & & 71.35 & 0.02 & & 71.21 & 0.02\\
%
% & 27707 & 71.08 & & 69.98 & 0.02 & & 69.77 & 0.02 & & 69.76 & 0.02\\
%
% & 27712 & 72.14 & & 71.29 & 0.01 & & 71.41 & 0.01 & & 71.53 & 0.01\\
%
% & 27713 & 69.38 & & 69.43 & 0.00 & & 69.53 & 0.00 & & 69.68 & 0.00\\
\end{table}

Table \ref{tabdescpt} summarizes a repeated sampling experiment
involving descriptive estimands at the zip code level.
For each of 100 simulation runs, we create $m=5$ synthetic data sets
using the observed mortality data (with $\tilde{Y}$) and the CART
synthesizers with $h
\in(1, 5, 10)$.
%Inferences are made using the methods of
%Reported statistics
%include the population values
%$Q$, the median of the $\bar{q}_5$ across the 100 simulation runs
%$ME$, the absolute percentage differences $PD=|(Q-median)/Q|$, and
%the coverage of synthetic $95\%$ confidence intervals $CS$. Three
%broad patterns can be seen in Table \ref{tabdescpt}. First, for
% For most estimands, the absolute
%percentage differences ($PD$) between the median of $\bar{q}_5$ and
%the observed point estimate are less than $10\%$.
For the percentage-related estimands,
the mean square error (MSE) is typically less than $3\%$, and for
age-related estimands, the MSE is typically less than 2.5 years.
%indicating that the synthetic data point estimates are close to their
%corresponding $Q$.
The MSEs for age-related
estimands are generally smaller than the other MSEs because age does
not vary
spatially as much as the other variables do;
hence, the synthesis process for age is comparatively robust to imperfect
modeling of the relationship between geographies and the attributes.
The MSEs tend to increase as $h$ increases, although the changes
for the most part are only 3\% or smaller.
%{\color{red} The $MSE$ for age-related
%estimands appears to be less than those of percentage-related
%estimands. This is possibly because the actual average age has less
%variability across different zip codes than the other two variables
%do, and hence the synthetic datasets shrink the age-related
%%estimands less than it does to the estimands of the two other
%variables.}
Overall, the results suggest that the synthetic data do
a reasonable job of preserving the aggregated spatial relationships
in the data for these variables.

%The two synthesis models produce
%significantly different estimation accuracy of ``$\%$ people with
%$E>14.5$''. The inferences from the model only simulating geography
%generally have larger $PD$ and smaller $CS$ than those from the
%model simulating both geography and non-geographic identifiers. Note
%that geography is imputed using all non-geographic variables in the
%first model but using all but $A$ and $R$ in the second model. The
%different inference performances on $E$ implies that different
%imputation models for geography may have different impacts on the
%marginal relationship between geography and non-geographic
%variables.Third, although $A$ and $R$ are unchanged in the first
%synthesis model but synthesized in the second model, the two models
%both produce quite reasonable and similar $PD$ and $CS$ on
%quantities involving $A$ and $R$.

We next evaluate inferences from two regression models. The first
is a~standard logistic regression of $\tilde{Y}$ on main effects for sex,
age and race. The second is a Bayesian spatial logistic regression
of $\tilde{Y}$ on main effects for sex, age and race that uses an
exponential covariance function for spatial random effects, as in
(\ref{eqspatial}). To aid in the evaluation of the
synthetic data sets, we randomly choose 2,470 people as a training set
to fit the models and the remaining 200 people as a
testing set to evaluate the predictive performance.
Because the sample size of this training set is large for fitting
hierarchical spatial random-effects models, we use Gaussian
predictive process models [\citet{BanerjeeEtAl08}] to reduce
computational burden. To do so, we select 100 knots by randomly
choosing a subset of the locations in the training set. We assign
flat prior distributions on regression coefficients $\bfb$, an
inverse Gamma $(2,1)$ prior for $\sigma^2_e$ and a uniform prior on
$(0.01,1)$ for $\phi_e$.
%The support for $\phi_e$ represents about 3-300 distance units for
%effective spatial ranges.
The same training sample, testing sample
and knots are used for all analyses, that is, we do not perform a
repeated sampling experiment because of computational burden of
estimating the spatial regression model. All models are estimated using
the ``spGLM''
function in R.

%
%t5 #&#
\begin{table}
\caption{Summary results for spatial and nonspatial logistic
regressions. Results include point and variance
estimates for regression coefficients, and misclassification rates (MR)}
\label{tabGLMGeoNongeo}
\begin{tabular*}{\tablewidth}{@{\extracolsep{\fill}}ld{2.2}cd{2.2}cd{2.2}cd{2.2}c@{}}
\hline
& \multicolumn{2}{c}{\textbf{Real data}} &
\multicolumn{2}{c}{$\bolds{h=1}$} &
\multicolumn{2}{c}{$\bolds{h=5}$} & \multicolumn{2}{c@{}}{$\bolds{h=10}$}\\
[-4pt]
& \multicolumn{2}{c}{\hspace*{-1.5pt}\hrulefill} & \multicolumn{2}{c}{\hspace*{-1.5pt}\hrulefill}
& \multicolumn{2}{c}{\hspace*{-1.5pt}\hrulefill} & \multicolumn{2}{c@{}}{\hspace*{-1.5pt}\hrulefill}\\
& \multicolumn{1}{c}{$\bolds{Q}$} & \multicolumn{1}{c}{$\bolds{\sqrt{T}}$}
& \multicolumn{1}{c}{$\bolds{\bar{q}_5}$} & \multicolumn{1}{c}{$\bolds{\sqrt{T_5}}$}
& \multicolumn{1}{c}{$\bolds{\bar{q}_5}$} & \multicolumn{1}{c}{$\bolds{\sqrt{T_5}}$}
& \multicolumn{1}{c}{$\bolds{\bar{q}_5}$} & \multicolumn{1}{c@{}}{$\bolds{\sqrt{T_5}}$}\\
\hline
\multicolumn{9}{@{}c@{}}{\textit{Nonspatial GLM}}\\[4pt]
Intercept & -0.85 & 0.18 & -0.76 & 0.21 & -0.71 & 0.20 & -0.67 &
0.19\\
Sex & 0.60 & 0.08 & 0.61 & 0.09 & 0.61 & 0.09 & 0.61 & 0.08\\
Race & 0.59 & 0.09 & 0.48 & 0.18 & 0.48 & 0.13 & 0.42 & 0.11\\
$\mbox{Age}\times100$ & 0.52 & 0.24 & 0.43 & 0.27 & 0.36 & 0.28 & 0.34 &
0.26\\
\hline
& \multicolumn{1}{c}{\textbf{MR}} & &\multicolumn{1}{c}{\textbf{MR}}
& & \multicolumn{1}{c}{\textbf{MR}} & & \multicolumn{1}{c@{}}{\textbf{MR}} \\
\hline
In-sample & 0.42 & & 0.42 & & 0.42 & & 0.42 & \\
Out-of-sample & 0.46 & & 0.47 & & 0.47 & & 0.47 & \\
[4pt]
\multicolumn{9}{@{}c@{}}{\textit{Spatial GLM}}\\
[4pt]
Intercept & -1.15 & 0.43 & -0.91 & 0.37 & -0.83 & 0.44 & -0.97 &
0.33\\
Sex & 0.74 & 0.10 & 0.64 & 0.09 & 0.67 & 0.09 & 0.68 & 0.10\\
Race & 0.82 & 0.12 & 0.62 & 0.23 & 0.61 & 0.15 & 0.56 & 0.13\\
$\mbox{Age}\times100$ & 0.68 & 0.25 & 0.65 & 0.31 & 0.48 & 0.28 & 0.53 &
0.28\\[2pt]
$\sigma_e^2$ & 1.83 & 0.96 & 1.82 & 1.20 & 1.31 & 0.73 & 1.13 &
0.53\\
$\phi_e$ & 0.05 & 0.02 & 0.06 & 0.02 & 0.06 & 0.01 & 0.06 & 0.01\\
\hline
& \multicolumn{1}{c}{\textbf{MR}} & &\multicolumn{1}{c}{\textbf{MR}}
& & \multicolumn{1}{c}{\textbf{MR}} & & \multicolumn{1}{c@{}}{\textbf{MR}} \\
\hline
In-sample & 0.22 & & 0.26 & & 0.25 & & 0.27 & \\
Out-of-sample & 0.32 & & 0.35 & & 0.31 & & 0.32 & \\
\hline
\end{tabular*}
\end{table}

Table \ref{tabGLMGeoNongeo} summarizes the original and synthetic data
inferences and predictions. For standard logistic regression, we
estimate the coefficients using the methods of
\citet{reiterpartsyn}. Misclassification rates are based on
predicting $\tilde{Y}_i=1$ when $p_i = 1/(1+e^{-\mathbf{x}'_i \bar
{\bfb}_5}) > 0.5$
and predicting $\tilde{Y}_i=0$ otherwise, where $\bar{\bfb}_5$ is
the vector
of synthetic point estimates for the coefficients. For the Bayesian
spatial logistic regression, we mix the posterior samples of the
coefficients from each of the five synthetic data sets, and report
the posterior mean and variance of the mixed samples.
% by their posterior
%means $\bar{q}_5$ and standard deviations $\sqrt{T_5}$ using the
%combined posterior samples from 5 synthetic datasets.
Misclassification rates are based on predicting $\tilde{Y}_i=1$ when the
posterior mean of $p_i$ across the five synthetic data sets exceeds
$0.5$ and predicting $\tilde{Y}_i=0$ otherwise. For both models, we compute
the in-sample misclassification rates as the proportions of
misclassified cases conditioned on the training set, and the
out-of-sample misclassification rates as the proportions of
misclassified cases conditioned on the test set. All out-of-sample
predictions for the Bayesian spatial logistic regression are carried
out using the ``spPredict'' function in R.

For the logistic regression, Table \ref{tabGLMGeoNongeo} indicates
that synthetic point estimates are generally close to those for the
observed data, although there is attenuation in the coefficients for
the synthesized variables. This attenuation increases with $h$.
Both in-sample and out-of-sample misclassification rates for the
synthetic data are similar to those for the observed data.

For the spatial regression, Table \ref{tabGLMGeoNongeo} indicates
that the synthetic point estimates are generally close to the
observed data estimates, again with increasing attenuation as $h$
gets large.
%within one standard error
%of the observed data point estimates.
%This implare substantial overlaps in
%the credible intervals of the synthetic estimates of regression
%coefficients $\beta$,
The spatial random effects parameters~$\sigma^2_e$ and~$\phi_e$
in the synthetic data are similar
to those from the observed data when $h=1$, but $\sigma^2_e$
declines toward zero as $h$ gets large. This indicates that large
values of $h$ can weaken the spatial associations in the synthetic data.

%Larger differences of
%coefficients for $R$ exist when $(\lambda,\phi,R,A)$ are replaced
%than when only $(\lambda,\phi)$ are replaced. For this study, it
%seems unclear about the impact of $h$ on both the inferences of
%coefficients and the misclassification errors.

It is also informative to compare the misclassification rates for
the spatial logistic regression in the synthetic data with the rates
for the nonspatial logistic regression in the observed data. In
particular, both in-sample and out-of-sample misclassification rates
are significantly lower in spatial logistic regression for the synthetic
data than those in nonspatial logistic regression for the
observed data. This suggests that, when spatial dependencies are
strong, releasing simulated geographies enables better predictions than
suppressing geography, even when race and sex are also simulated.

\begin{figure}

\includegraphics{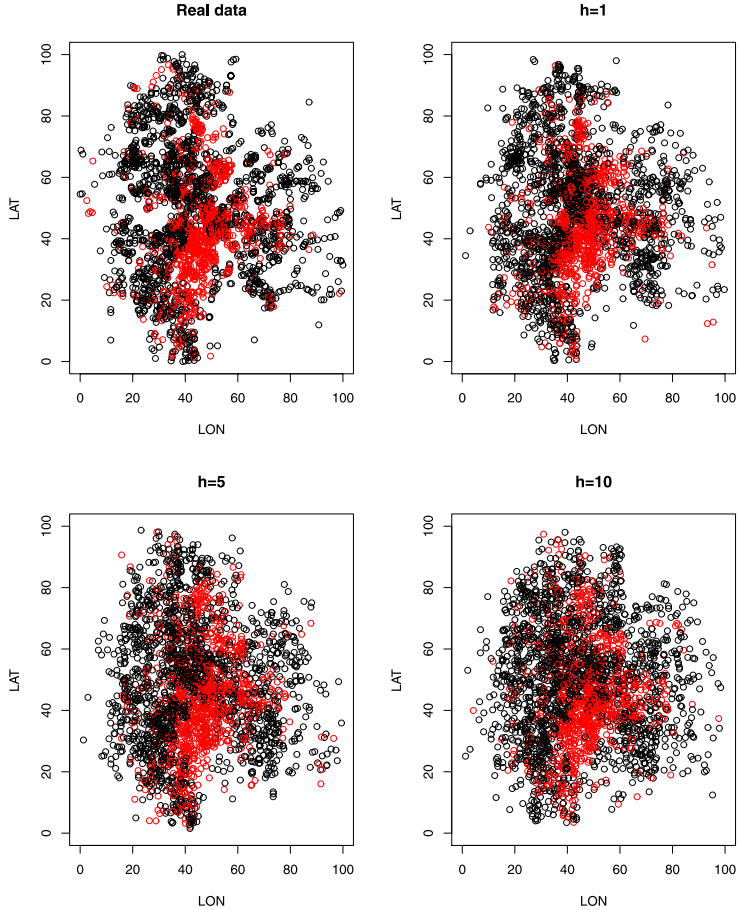}

\caption{Plots of the observed data (upper left) and synthetic data
sets for three levels of~$h$.
Red dots indicate locations of black people, and black dots indicate
locations of white people. All values of
$(\lambda,\phi,R,A)$ are synthesized.}
\label{fig1}
\vspace*{-3pt}
\end{figure}

The online supplement [\citet{WangReiter2011Supp}] reports the
results of the descriptive analyses and the
spatial regressions based on synthetic data sets generated from the
actual cause of death $Y$, which does
not exhibit strong spatial dependence.
% and synthetic data sets
%generated by another artificial $Y$ from a different logit model.
%eports these results. As is
%expected,
The results for the descriptive estimands are similar to, and even
slightly better than,
those from Table \ref{tabdescpt}. For the spatial regressions, the
synthetic data sets
appropriately reflect the lack of spatial dependence in~$Y$.
As a final illustration of the usefulness of the synthetic data sets,
Figure~\ref{fig1} displays maps of location by race for the actual
data and for three synthetic data sets ($m=1$) based on a CART
%
%f2 #&#
synthesizer with $h \in(1, 5, 10)$. Across all values of $h$, the
synthetic data sets preserve the spatial distribution of race
reasonably well.
%Of course, such maps present
%only one dimension of the quality of the synthetic datasets. Hence,
%in this section, we evaluate synthetic data inferences for a variety
%of estimands.
%All synthetic datasets are generated
%using on methods in Section \ref{MortalityGeneration}.

%s3.4 #&#
\subsection{Comparison against random noise addition}
When considering the merits of synthetic data approaches,
another relevant comparison is against other
disclosure limitation procedures rather than against the original
data, which cannot be made publicly available. We now compare the
synthetic data sets with only geography simulated against adding
random noise to geography, that is, moving an observed location to
another randomly drawn location. To make results comparable, we
perturb each $\mathbf{s}_i$ by drawing a~random
value~$\mathbf{s}_i^*$ from a bivariate normal distribution with a mean equal
to~$\mathbf{s}_i$ and a diagonal covariance matrix with standard deviations
equally set to be the corresponding ${R}_{1,i}/\sqrt{2}$.
Here, ${R}_{1,i}$ is computed assuming that, in the high-risk
scenario, only geography is synthesized and that $h=1$. In this way,
the synthetic and noise-infused data sets have roughly the same
${R}_1$ risks, because $\|\mathbf{s}^*_i-\mathbf{s}_i\|_2^2
\sim{1\over2} {R}_{1,i}^2 \chi^2_2$, and, hence,
$E(\|\mathbf{s}_i^*-\mathbf{s}_i\|_2^2) = R^2_{1,i}$. For
comparisons, we repeat
the analyses from Tables \ref{tabdescpt} and
\ref{tabGLMGeoNongeo}.

For the repeated sampling experiment, we add random noise to each
location independently 100 times, thus creating 100 noise-infused
data sets. For the noise infusion, four of the fourteen
percentage-related estimands in Table \ref{tabdescpt} have
$\mbox{MSE}>3\%$. In contrast, when synthesizing geography only with $h=1$,
%(as well
%as in Table \ref{tabdescpt} when additionally synthesizing age and
%race),
none of the percentage-related estimands have $\mbox{MSE}
> 3\%$; these results are reported in the online supplement
[\citet{WangReiter2011Supp}]. For the\vadjust{\goodbreak}
age-related estimands, the MSEs are similar for synthetic and
noise-infused data sets. Thus, for
comparable levels of disclosure risks, adding random
noise reduces the quality of inferences for the descriptive
estimands relative to synthetic data.

For the regression analyses, we estimate the Bayesian spatial
regression with one data set generated by adding random
noise to geography only. The in-sample and out-of-sample
misclassification rates are 0.32 and 0.38, respectively, for this
noise-infused data set. We observed similar misclassification rates when
repeating this analysis three more times. These misclassification
rates are substantially larger than those for the corresponding
synthetic data sets reported in the supplement (as well as those in
Table \ref{tabGLMGeoNongeo}), again suggesting that, for comparable risk
levels, random noise does not preserve spatial relationships as well
as synthetic data.\vadjust{\goodbreak}

%s4 #&#
\section{Concluding remarks}\label{discussion}
%Current methods of protecting geographic variables create problems
%for inferences, including ecological fallacies, lack of fine spatial
%resolution, and even complete obliteration of spatial dependencies.
%This research has developed improved alternatives to current methods
%that have the potential to preserve spatial dependencies without
%compromising confidentiality. Better methods of data sharing can
%lead to better analysis of secondary data, and hence better science.
%%Ultimately, by enhancing the ability of scientists to exploit finely
%resolved geographic data, these proposed new methods have generate
%new insights that may improve clinical care and practice, as well as
%health more generally.

Although synthesizing geographies via modeling, such
as the CART approach here, can preserve some spatial analyses, it
does not preserve all of them. For example, two records close in
space in the original data will not necessarily be close in space in
the synthetic data, because their locations are independently
generated from the response distribution.
%Such movements could be
%necessary to protect confidentiality.
Additionally, simulated
geographies may not preserve analyses when used to link the
synthetic data with other data containing geography, since the
simulated locations are conditionally independent of the variables
in the linked data set that are not included in the synthesis model.
Evaluating the impacts of synthetic geographies on linked analysis
is a future extension of this research.

When synthesizing the nongeographic quasi-identifiers, we
controlled for location as predictors in the model. An alternative
approach is to
simulate from hierarchical spatial models for point-referenced data,
or perhaps from area-level models by aggregating locations
[\citet{BGC2004}]. With large data sets, fitting spatial random
effects models can be computationally challenging, although this can
be overcome using approximations from the spatial statistics
literature. Another strategy is to mask attribute data using spatial
smoothing techniques [\citet{ZhouEtAl2010}]. We note that applying
either of these approaches alone, that is, without simulating geography,
leaves the original fine geography on the file, which may be too high
of a disclosure risk. Evaluating the potential gains in disclosure risk and
data usefulness of such strategies over the simple CART synthesizer
for attributes utilized here is an area open for further theoretical
and empirical investigation.

To implement the CART synthesizer, data stewards need to select the
tuning parameters of the trees, that is,
the minimum number of observations per leaf and the splitting criteria.
These parameters control the size of the tree: increasing them
results in smaller trees, and decreasing them results in larger trees.
Based on our experience, we recommend that data
stewards begin by setting the minimum deviance in the splitting
criteria to a small number, like 0.0001 or even smaller, and requiring
at least five records per leaf. These are
typical default values for many applications and software routines for
regression trees. The data steward then evaluates the disclosure
risk and data utility associated with the synthetic data sets. If the
risks are too high, the data steward can re-tune the parameters
for the variables that are not sufficiently altered by the synthesis to
grow smaller trees for those variables
[\citet{reitercart}].
We did not prune the leaves further, as experiments with further
pruning worsened the quality of the synthetic data sets without substantially
improving the confidentiality protection. Growing larger trees can
increase the quality of the synthetic data sets. However, it
increases the time to run the synthesizer. Further, it can increase
disclosure risks, for example, using trees with one observation per
leaf reproduces the original data.\vadjust{\goodbreak}

The CART synthesizer has appealing features: it handles continuous,
categorical and mixed data; captures nonlinear relationships and
complex interactions automatically; and runs quickly on large data sets.
However, CART synthesizers can run into computational difficulties when
categorical
variables have many (e.g., $>$20) levels. Additionally, when some levels
have low incidence rates in the data, the CART synthesizer can have
difficulty preserving relationships involving those levels
[\citet{reitercart}].

For simulation purposes, we illustrated the CART synthesizer using
only $n=2\mbox{,}670$ records. This facilitated estimation of the spatial
regressions with each of the resulting synthetic data sets. In
extended investigations, we found that the CART synthesis process
readily scaled for tens of thousands of mortality records. Other
applications using CART synthesizers for nongeographic attributes
[\citet{reiterisr}, \citet{joergjas}] indicate that it can
be applied in
surveys of dimensions typical of many government surveys. When data
stewards need to synthesize locations for a very large number, for example,
millions, of records, a computationally convenient strategy is to
partition the data into geographical strata of manageable size (tens
of thousands of records), and simulate latitudes and longitudes (and
attributes) by running the synthesizer independently within each
stratum.

\begin{supplement}%[id=suppA]
\stitle{Computational details and further results\\}
\slink[doi,text={10.1214/11-AOAS506SUPP}]{10.1214/11-AOAS506SUPP} %[doi,text={...}] - jei reikia
%suskaldyti doi
\slink[url]{http://lib.stat.cmu.edu/aoas/506/supplement.pdf}
\sdatatype{.pdf}
\sdescription{Computational details for geography
disclosure and identification risks in Sections
\ref{SectionRiskAttri} and \ref{SectionRiskID}; further
analytical validity results; and results based on genuine cause of
death.}
\end{supplement}

% imsref loaded by lrinkeviciute, 2011-12-01 08:44:52
% imsref loaded by lrinkeviciute, 2011-12-01 08:51:59
% imsref loaded by lrinkeviciute, 2011-12-01 09:02:58
%

\printaddresses


\begin{thebibliography}{41}
% BibTex style file: ims.bst, 2011-05-30
% Default style options (sort=0,type=number).
% Used options (sort=1,type=nameyear).

%b1 #&#
\bibitem[\protect\citeauthoryear{Armstrong, Rushton and
Zimmerman}{1999}]{rushton}
%
\begin{barticle}[author]
\bauthor{\bsnm{Armstrong},~\bfnm{M.~P.}\binits{M.~P.}},
\bauthor{\bsnm{Rushton},~\bfnm{G.}\binits{G.}} \AND
\bauthor{\bsnm{Zimmerman},~\bfnm{D.~L.}\binits{D.~L.}}
(\byear{1999}).
\btitle{Geographically masking health data to preserve confidentiality}.
\bjournal{Stat. Med.}
\bvolume{18}
\bpages{495--525}.
\bptok{imsref}%
\end{barticle}
%
\endbibitem

%b2 #&#
\bibitem[\protect\citeauthoryear{Banerjee, Gelfand and
Carlin}{2004}]{BGC2004}
%
\begin{bbook}[author]
\bauthor{\bsnm{Banerjee},~\bfnm{Sudipto}\binits{S.}},
\bauthor{\bsnm{Gelfand},~\bfnm{Alan~E.}\binits{A.~E.}} \AND
\bauthor{\bsnm{Carlin},~\bfnm{Bradley~P.}\binits{B.~P.}}
(\byear{2004}).
\btitle{Hierarchical Modeling and Analysis for Spatial Data}.
\bpublisher{Chapman and Hall/CRC}, \baddress{Boca Raton, FL}.
\bptok{imsref}%
\end{bbook}
%
\endbibitem

%b3 #&#
\bibitem[\protect\citeauthoryear{Banerjee et~al.}{2008}]{BanerjeeEtAl08}
%
\begin{barticle}[mr]
\bauthor{\bsnm{Banerjee},~\bfnm{Sudipto}\binits{S.}},
\bauthor{\bsnm{Gelfand},~\bfnm{Alan~E.}\binits{A.~E.}},
\bauthor{\bsnm{Finley},~\bfnm{Andrew~O.}\binits{A.~O.}} \AND
\bauthor{\bsnm{Sang},~\bfnm{Huiyan}\binits{H.}}
(\byear{2008}).
\btitle{Gaussian predictive process models for large spatial data sets}.
\bjournal{J. R. Stat. Soc. Ser. B Stat. Methodol.}
\bvolume{70}
\bpages{825--848}.
\bid{doi={10.1111/j.1467-9868.2008.00663.x}, issn={1369-7412}, mr={2523906}}
\bptok{imsref}%
\end{barticle}
%
\endbibitem

%b4 #&#
\bibitem[\protect\citeauthoryear{Breiman et~al.}{1984}]{breiman1984}
%
\begin{bbook}[author]
\bauthor{\bsnm{Breiman},~\bfnm{L.}\binits{L.}},
\bauthor{\bsnm{Friedman},~\bfnm{J.~H.}\binits{J.~H.}},
\bauthor{\bsnm{Olshen},~\bfnm{R.~A.}\binits{R.~A.}} \AND
\bauthor{\bsnm{Stone},~\bfnm{C.~J.}\binits{C.~J.}}
(\byear{1984}).
\btitle{Classification and Regression Trees}.
\bpublisher{Wadsworth}, \baddress{Belmont}.
\bptok{imsref}%
\end{bbook}
%
\endbibitem

%b5 #&#
\bibitem[\protect\citeauthoryear{Caiola and Reiter}{2010}]{reitcaiola}
%
\begin{barticle}[mr]
\bauthor{\bsnm{Caiola},~\bfnm{Gregory}\binits{G.}} \AND
\bauthor{\bsnm{Reiter},~\bfnm{Jerome~P.}\binits{J.~P.}}
(\byear{2010}).
\btitle{Random forests for generating partially synthetic, categorical data}.
\bjournal{Trans. Data Priv.}
\bvolume{3}
\bpages{27--42}.
\bid{issn={1888-5063}, mr={2725418}}
\bptok{imsref}%
\end{barticle}
%
\endbibitem

%b6 #&#
\bibitem[\protect\citeauthoryear{Chipman, George and
McCulloch}{2010}]{Chipman10BART}
%
\begin{barticle}[mr]
\bauthor{\bsnm{Chipman},~\bfnm{Hugh~A.}\binits{H.~A.}},
\bauthor{\bsnm{George},~\bfnm{Edward~I.}\binits{E.~I.}} \AND
\bauthor{\bsnm{McCulloch},~\bfnm{Robert~E.}\binits{R.~E.}}
(\byear{2010}).
\btitle{B{ART}: {B}ayesian additive regression trees}.
\bjournal{Ann. Appl. Stat.}
\bvolume{4}
\bpages{266--298}.
\bid{doi={10.1214/09-AOAS285}, issn={1932-6157}, mr={2758172}}
\bptok{imsref}%
\end{barticle}
%
\endbibitem

%b7 #&#
\bibitem[\protect\citeauthoryear{Dalenius and Reiss}{1982}]{dalenius}
%
\begin{barticle}[mr]
\bauthor{\bsnm{Dalenius},~\bfnm{Tore}\binits{T.}} \AND
\bauthor{\bsnm{Reiss},~\bfnm{Steven~P.}\binits{S.~P.}}
(\byear{1982}).
\btitle{Data-swapping: A technique for disclosure control}.
\bjournal{J. Statist. Plann. Inference}
\bvolume{6}
\bpages{73--85}.
\bid{doi={10.1016/0378-3758(82)90058-1}, issn={0378-3758}, mr={0653248}}
\bptok{imsref}%
\end{barticle}
%
\endbibitem

%b8 #&#
\bibitem[\protect\citeauthoryear{De'ath}{2002}]{Death08}
%
\begin{barticle}[author]
\bauthor{\bsnm{De'ath},~\bfnm{Glenn}\binits{G.}}
(\byear{2002}).
\btitle{Multivariate regression trees: A new technique for modeling species
environment relationships}.
\bjournal{Ecology}
\bvolume{83}
\bpages{1105--1117}.
\bptok{imsref}%
\end{barticle}
%
\endbibitem

%b9 #&#
\bibitem[\protect\citeauthoryear{Drechsler}{2011}]{joergjas}
%
\begin{barticle}[author]
\bauthor{\bsnm{Drechsler},~\bfnm{J.}\binits{J.}}
(\byear{2011}).
\btitle{New data dissemination approaches in old {E}urope---{S}ynthetic
datasets for a {G}erman establishment survey}.
\bjournal{J. Appl. Stat.}
\bnote{To appear}.
\bptok{imsref}%
\end{barticle}
%
\endbibitem

%b10 #&#
\bibitem[\protect\citeauthoryear{Drechsler and Reiter}{2008}]{drechreiter08}
%
\begin{bincollection}[author]
\bauthor{\bsnm{Drechsler},~\bfnm{J.}\binits{J.}} \AND
\bauthor{\bsnm{Reiter},~\bfnm{J.~P.}\binits{J.~P.}}
(\byear{2008}).
\btitle{Accounting for intruder uncertainty due to sampling when estimating
identification disclosure risks in partially synthetic data}.
In \bbooktitle{Privacy in Statistical Databases (LNCS 5262)}
(\beditor{\bfnm{J.}\binits{J.}~\bsnm{Domingo-Ferrer}} \AND
\beditor{\bfnm{Y.}\binits{Y.}~\bsnm{Saygin}}, eds.)
\bpages{227--238}.
\bpublisher{Springer}, \baddress{New York}.
\bptok{imsref}%
\end{bincollection}
%
\endbibitem

%b11 #&#
\bibitem[\protect\citeauthoryear{Drechsler and
Reiter}{2010}]{reitdrechcensus}
%
\begin{barticle}[mr]
\bauthor{\bsnm{Drechsler},~\bfnm{J{\"o}rg}\binits{J.}} \AND
\bauthor{\bsnm{Reiter},~\bfnm{Jerome~P.}\binits{J.~P.}}
(\byear{2010}).
\btitle{Sampling with synthesis: A new approach for releasing public
use census
microdata}.
\bjournal{J. Amer. Statist. Assoc.}
\bvolume{105}
\bpages{1347--1357}.
\bid{doi={10.1198/jasa.2010.ap09480}, issn={0162-1459}, mr={2796555}}
\bptok{imsref}%
\end{barticle}
%
\endbibitem

%b12 #&#
\bibitem[\protect\citeauthoryear{Duncan and Lambert}{1989}]{dl89}
%
\begin{barticle}[author]
\bauthor{\bsnm{Duncan},~\bfnm{G.~T.}\binits{G.~T.}} \AND
\bauthor{\bsnm{Lambert},~\bfnm{D.}\binits{D.}}
(\byear{1989}).
\btitle{The risk of disclosure for microdata}.
\bjournal{Journal of Business and Economic Statistics}
\bvolume{7}
\bpages{207--217}.
\bptok{imsref}%
\end{barticle}
%
\endbibitem

%b13 #&#
\bibitem[\protect\citeauthoryear{Federal Register}{2000}]{hipaa}
%
\begin{bmisc}[author]
\borganization{Federal Register}
(\byear{2000}).
\bhowpublished{Standards for privacy of individually identifiable health
information---Final privacy rule. 45 C. F. R. Parts 160 and 164, Dept. Health
and Human Services, Office of the Secretary, Washington, DC}.
\bptok{imsref}%
\end{bmisc}
%
\endbibitem

%b14 #&#
\bibitem[\protect\citeauthoryear{Fienberg, Makov and
Sanil}{1997}]{fms97}
%
\begin{barticle}[author]
\bauthor{\bsnm{Fienberg},~\bfnm{S.~E.}\binits{S.~E.}},
\bauthor{\bsnm{Makov},~\bfnm{U.~E.}\binits{U.~E.}} \AND
\bauthor{\bsnm{Sanil},~\bfnm{A.~P.}\binits{A.~P.}}
(\byear{1997}).
\btitle{A {B}ayesian approach to data disclosure: {O}ptimal intruder behavior
for continuous data}.
\bjournal{Journal of Official Statistics}
\bvolume{13}
\bpages{75--89}.
\bptok{imsref}%
\end{barticle}
%
\endbibitem

%b15 #&#
\bibitem[\protect\citeauthoryear{Fienberg and McIntyre}{2004}]{fienmac}
%
\begin{bincollection}[author]
\bauthor{\bsnm{Fienberg},~\bfnm{S.~E.}\binits{S.~E.}} \AND
\bauthor{\bsnm{McIntyre},~\bfnm{S.~E.}\binits{S.~E.}}
(\byear{2004}).
\btitle{Data swapping: Variations on a theme by Dalenius and Reese}.
In \bbooktitle{Privacy in Statistical Databases}
(\beditor{\bfnm{J.}\binits{J.}~\bsnm{Domingo-Ferrer}} \AND
\beditor{\bfnm{V.}\binits{V.}~\bsnm{Torra}}, eds.)
\bpages{14--29}.
\bpublisher{Springer}, \baddress{New York}.
\bptok{imsref}%
\end{bincollection}
%
\endbibitem

%b16 #&#
\bibitem[\protect\citeauthoryear{Freedman}{2004}]{freedman04}
%
\begin{bincollection}[author]
\bauthor{\bsnm{Freedman},~\bfnm{D.~A.}\binits{D.~A.}}
(\byear{2004}).
\btitle{The ecological fallacy}.
In \bbooktitle{Encyclopedia of Social Science Research Methods}
(\beditor{\bfnm{M.}\binits{M.}~\bsnm{Lewis-Beck}},
\beditor{\bfnm{A.}\binits{A.}~\bsnm{Bryman}} \AND
\beditor{\bfnm{T.~F.}\binits{T.~F.}~\bsnm{Liao}}, eds.)
\bvolume{1}
\bpages{293}.
\bpublisher{Sage}, \baddress{Thousand Oaks, CA}.
\bptok{imsref}%
\end{bincollection}
%
\endbibitem

%b17 #&#
\bibitem[\protect\citeauthoryear{Fuller}{1993}]{fuller1993}
%
\begin{barticle}[author]
\bauthor{\bsnm{Fuller},~\bfnm{W.~A.}\binits{W.~A.}}
(\byear{1993}).
\btitle{Masking procedures for microdata disclosure limitation}.
\bjournal{Journal of Official Statistics}
\bvolume{9}
\bpages{383--406}.
\bptok{imsref}%
\end{barticle}
%
\endbibitem

%b18 #&#
\bibitem[\protect\citeauthoryear{Gomatam et~al.}{2005}]{statsciserverkarr}
%
\begin{barticle}[mr]
\bauthor{\bsnm{Gomatam},~\bfnm{S.}\binits{S.}},
\bauthor{\bsnm{Karr},~\bfnm{A.~F.}\binits{A.~F.}},
\bauthor{\bsnm{Reiter},~\bfnm{J.~P.}\binits{J.~P.}} \AND
\bauthor{\bsnm{Sanil},~\bfnm{A.~P.}\binits{A.~P.}}
(\byear{2005}).
\btitle{Data dissemination and disclosure limitation in a world without
microdata: A risk-utility framework for remote access analysis servers}.
\bjournal{Statist. Sci.}
\bvolume{20}
\bpages{163--177}.
\bid{doi={10.1214/088342305000000043}, issn={0883-4237}, mr={2183447}}
\bptok{imsref}%
\end{barticle}
%
\endbibitem

%b19 #&#
\bibitem[\protect\citeauthoryear{Health and Retirement Study}{2007}]{hrs}
%
\begin{bmisc}[author]
\borganization{Health and Retirement Study}
(\byear{2007}).
\bhowpublished{Data Description and Usage (2006 Core, Early, Version~2.0).
Available at
\texttt{\href{http://hrsonline.isr.umich.edu/meta/2006/core/desc/h06dd.pdf}{http://hrsonline.isr.umich.edu/meta/2006/core/desc/}
\href{http://hrsonline.isr.umich.edu/meta/2006/core/desc/h06dd.pdf}{h06dd.pdf}}.}
\bptok{imsref}%
\end{bmisc}
%
\endbibitem

%b20 #&#
\bibitem[\protect\citeauthoryear{Kinney
et~al.}{2011}]{lbdisr}
%
\begin{bmisc}[author]
\bauthor{\bsnm{Kinney},~\bfnm{S.~K.}\binits{S.~K.}},
\bauthor{\bsnm{Reiter},~\bfnm{J.~P.}\binits{J.~P.}},
\bauthor{\bsnm{Reznek},~\bfnm{A.~P.}\binits{A.~P.}},
\bauthor{\bsnm{Miranda},~\bfnm{J.}\binits{J.}},
\bauthor{\bsnm{Jarmin},~\bfnm{R.~S.}\binits{R.~S.}} \AND
\bauthor{\bsnm{Abowd},~\bfnm{J.~M.}\binits{J.~M.}}
(\byear{2011}).
\bhowpublished{Towards unrestricted public use business microdata: {T}he
synthetic {L}ongitudinal {B}usiness {D}atabase. Technical report,
Center for
Economic Studies Working Paper CES-WP-11-04, Census Bureau, Washington, DC.}
\bptok{imsref}%
\end{bmisc}
%
\endbibitem

%b21 #&#
\bibitem[\protect\citeauthoryear{Little}{1993}]{little1993}
%
\begin{barticle}[author]
\bauthor{\bsnm{Little},~\bfnm{R.~J.~A.}\binits{R.~J.~A.}}
(\byear{1993}).
\btitle{Statistical analysis of masked data}.
\bjournal{Journal of Official Statistics}
\bvolume{9}
\bpages{407--426}.
\bptok{imsref}%
\end{barticle}
%
\endbibitem

%b39 #&#
\bibitem[\protect\citeauthoryear{Little, Liu and Raghunathan}{2004}]{litragliu}
%
\begin{bincollection}[author]
\bauthor{\bsnm{Little},~\bfnm{R.~J.~A.}\binits{R.~J.~A.}},
 \bauthor{\bsnm{Liu},~\bfnm{F.}\binits{F.}} \AND
 \bauthor{\bsnm{Raghunathan},~\bfnm{T.~E.}\binits{T.~E.}}
(\byear{2004}).
\btitle{Statistical disclosure techniques based on multiple imputation}.
In \bbooktitle{Applied Bayesian Modeling and Causal Inference from
 Incomplete-Data Perspectives}
(\beditor{\bfnm{A.}\binits{A.}~\bsnm{Gelman}} \AND
 \beditor{\bfnm{X.~L.}\binits{X.~L.}~\bsnm{Meng}}, eds.)
\bpages{141--152}.
\bpublisher{Wiley}, \baddress{New York}.
%Edited by
% Andrew Gelman and Xiao-Li Meng}.
\bptok{imsref}%
\end{bincollection}
%
\endbibitem

%b22 #&#
\bibitem[\protect\citeauthoryear{Machanavajjhala et~al.}{2008}]{onthemap}
%
\begin{binproceedings}[author]
\bauthor{\bsnm{Machanavajjhala},~\bfnm{A.}\binits{A.}},
\bauthor{\bsnm{Kifer},~\bfnm{D.}\binits{D.}},
\bauthor{\bsnm{Abowd},~\bfnm{J.}\binits{J.}},
\bauthor{\bsnm{Gehrke},~\bfnm{J.}\binits{J.}} \AND
\bauthor{\bsnm{Vilhuber},~\bfnm{L.}\binits{L.}}
(\byear{2008}).
\btitle{Privacy: {T}heory meets practice on the map}.
In \bbooktitle{IEEE 24th International Conference on Data Engineering}
\bpages{277--286}.
\bptok{imsref}%
\end{binproceedings}
%
\endbibitem

%b23 #&#
\bibitem[\protect\citeauthoryear{National Research Council}{2005}]{nrc05}
%
\begin{bmisc}[author]
\borganization{National Research Council}
(\byear{2005}).
\bhowpublished{Expanding access to research data: Reconciling risks and
opportunities. Panel on Data Access for Research Purposes, Committee on
National Statistics, Division of Behavioral and Social Sciences and
Education. The National Academies Press, Washington, DC.}
\bptok{imsref}%
\end{bmisc}
%
\endbibitem

%b24 #&#
\bibitem[\protect\citeauthoryear{National Research Council}{2007}]{nrc07}
%
\begin{bmisc}[author]
\borganization{National Research Council}
(\byear{2007}).
\bhowpublished{Putting people on the map: Protecting confidentiality with
linked social-spatial data. Panel on Confidentiality Issues Arising
from the
Integration of Remotely Sensed and Self-Identifying Data, Committee on the
Human Dimensions of Global Change, Division of Behavioral and Social Sciences
and Education. The National Academies Press, Washington, DC.}
\bptok{imsref}%
\end{bmisc}
%
\endbibitem

%b25 #&#
\bibitem[\protect\citeauthoryear{Reiter}{2003}]{reiterpartsyn}
%
\begin{barticle}[author]
\bauthor{\bsnm{Reiter},~\bfnm{J.~P.}\binits{J.~P.}}
(\byear{2003}).
\btitle{Inference for partially synthetic, public use microdata sets}.
\bjournal{Survey Methodology}
\bvolume{29}
\bpages{181--189}.
\bptok{imsref}%
\end{barticle}
%
\endbibitem

%b26 #&#
\bibitem[\protect\citeauthoryear{Reiter}{2004a}]{reiterchance}
%
\begin{barticle}[mr]
\bauthor{\bsnm{Reiter},~\bfnm{Jerome~P.}\binits{J.~P.}}
(\byear{2004}a).
\btitle{New approaches to data dissemination: A glimpse into the
future (?)}.
\bjournal{Chance}
\bvolume{17}
\bpages{11--15}.
\bid{issn={0933-2480}, mr={2061931}}
\bptok{imsref}%
\end{barticle}
%
\endbibitem

%b27 #&#
\bibitem[\protect\citeauthoryear{Reiter}{2004b}]{reitermi}
%
\begin{barticle}[author]
\bauthor{\bsnm{Reiter},~\bfnm{J.~P.}\binits{J.~P.}}
(\byear{2004}b).
\btitle{Simultaneous use of multiple imputation for missing data and disclosure
limitation}.
\bjournal{Survey Methodology}
\bvolume{30}
\bpages{235--242}.
\bptok{imsref}%
\end{barticle}
%
\endbibitem

%b28 #&#
\bibitem[\protect\citeauthoryear{Reiter}{2005a}]{reiter05}
%
\begin{barticle}[author]
\bauthor{\bsnm{Reiter},~\bfnm{J.~P.}\binits{J.~P.}}
(\byear{2005}a).
\btitle{Estimating identification risks in microdata}.
\bjournal{J. Amer. Statist. Assoc.}
\bvolume{100}
\bpages{1103--1113}.
\bptok{imsref}%
\end{barticle}
%
\endbibitem

%b29 #&#
\bibitem[\protect\citeauthoryear{Reiter}{2005b}]{reiter2002a}
%
\begin{barticle}[mr]
\bauthor{\bsnm{Reiter},~\bfnm{Jerome~P.}\binits{J.~P.}}
(\byear{2005}b).
\btitle{Releasing multiply imputed, synthetic public use microdata: An
illustration and empirical study}.
\bjournal{J. Roy. Statist. Soc. Ser. A}
\bvolume{168}
\bpages{185--205}.
\bid{doi={10.1111/j.1467-985X.2004.00343.x}, issn={0964-1998}, mr={2113234}}
\bptok{imsref}%
\end{barticle}
%
\endbibitem

%b30 #&#
\bibitem[\protect\citeauthoryear{Reiter}{2005c}]{reitermulti}
%
\begin{barticle}[mr]
\bauthor{\bsnm{Reiter},~\bfnm{J.~P.}\binits{J.~P.}}
(\byear{2005}c).
\btitle{Significance tests for multi-component estimands from multiply imputed,
synthetic microdata}.
\bjournal{J. Statist. Plann. Inference}
\bvolume{131}
\bpages{365--377}.
\bid{doi={10.1016/j.jspi.2004.02.003}, issn={0378-3758}, mr={2139378}}
\bptok{imsref}%
\end{barticle}
%
\endbibitem

%b31 #&#
\bibitem[\protect\citeauthoryear{Reiter}{2005d}]{reitercart}
%
\begin{barticle}[author]
\bauthor{\bsnm{Reiter},~\bfnm{J.~P.}\binits{J.~P.}}
(\byear{2005}d).
\btitle{Using {CART} to generate partially synthetic, public use microdata}.
\bjournal{Journal of Official Statistics}
\bvolume{21}
\bpages{441--462}.
\bptok{imsref}%
\end{barticle}
%
\endbibitem

%b32 #&#
\bibitem[\protect\citeauthoryear{Reiter}{2009}]{reiterisr}
%
\begin{barticle}[author]
\bauthor{\bsnm{Reiter},~\bfnm{J.~P.}\binits{J.~P.}}
(\byear{2009}).
\btitle{Using multiple imputation to integrate and disseminate confidential
microdata}.
\bjournal{International Statistical Review}
\bvolume{77}
\bpages{179--195}.
\bptok{imsref}%
\end{barticle}
%
\endbibitem

%b33 #&#
\bibitem[\protect\citeauthoryear{Reiter and Mitra}{2009}]{reitermitra07}
%
\begin{barticle}[author]
\bauthor{\bsnm{Reiter},~\bfnm{J.~P.}\binits{J.~P.}} \AND
\bauthor{\bsnm{Mitra},~\bfnm{R.}\binits{R.}}
(\byear{2009}).
\btitle{Estimating risks of identification disclosure in partially synthetic
data}.
\bjournal{Journal of Privacy and Confidentiality}
\bvolume{1}
\bpages{99--110}.
\bptok{imsref}%
\end{barticle}
%
\endbibitem

%b34 #&#
\bibitem[\protect\citeauthoryear{Robinson}{1950}]{ecologicalinference}
%
\begin{barticle}[author]
\bauthor{\bsnm{Robinson},~\bfnm{W.~S.}\binits{W.~S.}}
(\byear{1950}).
\btitle{Ecological correlations and the behavior of individuals}.
\bjournal{American Sociological Review}
\bvolume{15}
\bpages{351--357}.
\bptok{imsref}%
\end{barticle}
%
\endbibitem

%b35 #&#
\bibitem[\protect\citeauthoryear{Rubin}{1981}]{rubin1981}
%
\begin{barticle}[mr]
\bauthor{\bsnm{Rubin},~\bfnm{Donald~B.}\binits{D.~B.}}
(\byear{1981}).
\btitle{The {B}ayesian bootstrap}.
\bjournal{Ann. Statist.}
\bvolume{9}
\bpages{130--134}.
\bid{issn={0090-5364}, mr={0600538}}
\bptok{imsref}%
\end{barticle}
%
\endbibitem

%b36 #&#
\bibitem[\protect\citeauthoryear{Sweeney}{2001}]{sweeney2001}
%
\begin{bmisc}[author]
\bauthor{\bsnm{Sweeney},~\bfnm{L.~A.}\binits{L.~A.}}
(\byear{2001}).
\bhowpublished{Computational disclosure control: A primer on data privacy
protection. Ph.D. thesis, MIT, Cambridge, MA.}
\bptok{imsref}%
\end{bmisc}
%
\endbibitem

%b37 #&#
\bibitem[\protect\citeauthoryear{Van{W}ey et~al.}{2005}]{guttman}
%
\begin{barticle}[author]
\bauthor{\bsnm{Van{W}ey},~\bfnm{L.~K.}\binits{L.~K.}},
\bauthor{\bsnm{Rindfuss},~\bfnm{R.~R.}\binits{R.~R.}},
\bauthor{\bsnm{Guttman},~\bfnm{M.~P.}\binits{M.~P.}},
\bauthor{\bsnm{Entwisle},~\bfnm{B.}\binits{B.}} \AND
\bauthor{\bsnm{Balk},~\bfnm{D.~L.}\binits{D.~L.}}
(\byear{2005}).
\btitle{Confidentiality and spatially explicit data: Concerns and challenges}.
\bjournal{Proc. Natl. Acad. Sci. USA}
\bvolume{102}
\bpages{15337--15342}.
\bptok{imsref}%
\end{barticle}
%
\endbibitem

%b38 #&#
\bibitem[\protect\citeauthoryear{Wang and Reiter}{2011}]{WangReiter2011Supp}
%
\begin{bmisc}[author]
\bauthor{\bsnm{Wang},~\bfnm{Hao}\binits{H.}} \AND
\bauthor{\bsnm{Reiter},~\bfnm{Jerome}\binits{J.}}
(\byear{2011}).
\bhowpublished{Supplement to ``Multiple imputation for sharing precise
geographies in public use data.''
\href{http://dx.doi.org/10.1214/11-AOAS506SUPP}{DOI:10.1214/11-AOAS506SUPP}.}
\bptok{imsref}%
\end{bmisc}
%
\endbibitem

%b40 #&#
\bibitem[\protect\citeauthoryear{Zhou, Dominici and
Louis}{2010}]{ZhouEtAl2010}
%
\begin{barticle}[mr]
\bauthor{\bsnm{Zhou},~\bfnm{Yijie}\binits{Y.}},
\bauthor{\bsnm{Dominici},~\bfnm{Francesca}\binits{F.}} \AND
\bauthor{\bsnm{Louis},~\bfnm{Thomas~A.}\binits{T.~A.}}
(\byear{2010}).
\btitle{A smoothing approach for masking spatial data}.
\bjournal{Ann. Appl. Stat.}
\bvolume{4}
\bpages{1451--1475}.
\bid{doi={10.1214/09-AOAS325}, issn={1932-6157}, mr={2758336}}
\bptok{imsref}%
\end{barticle}
%
\endbibitem

\end{thebibliography}
\end{document}